%% file: ckt-GSE.tex
\documentclass[journal]{IEEEtran}
\IEEEoverridecommandlockouts
\usepackage{cite}
\usepackage{amsmath,amssymb,amsfonts}
\usepackage{algorithmic}
\usepackage{graphicx}
\usepackage{textcomp}
\usepackage{xcolor}

\usepackage{tabularx}

\usepackage{mdwlist}
\usepackage[ruled,vlined,shortend, linesnumbered]{algorithm2e} 
\usepackage[]{algorithmic} 
\usepackage{booktabs}
\usepackage{color}

\usepackage{amsthm,amssymb}
\usepackage{amsmath}
\usepackage[]{caption}
\usepackage{subcaption}
\usepackage{dblfloatfix} 

\usepackage{booktabs}

\usepackage{multirow}
\usepackage{mathrsfs}
\usepackage{amsbsy}
\usepackage[mathscr]{eucal} 
\usepackage[flushleft]{threeparttable}
\usepackage[hidelinks]{hyperref}
\usepackage{comment}
\usepackage{paralist}
\usepackage{pifont}

\usepackage{bbding} 

\def\BibTeX{{\rm B\kern-.05em{\sc i\kern-.025em b}\kern-.08em
    T\kern-.1667em\lower.7ex\hbox{E}\kern-.125emX}}

\usepackage[colorinlistoftodos]{todonotes}
\usepackage{xcolor}

\theoremstyle{definition}
\newtheorem{definition}{Definition}[section]

\usepackage{amsthm}
\usepackage{amsmath}
\usepackage{amssymb}
\usepackage{thmtools}
\usepackage{thm-restate}
\usepackage{hyperref}

\usepackage{mathtools}

\newcommand{\cmark}{\ding{52}}%
\newcommand{\xmark}{\ding{56}}%

\newcommand{\statusmeass}{switch status data}
\newcommand{\contmeass}{continuous measurements}

\newcommand{\statusests}{switch statuses}

\newcommand{\statusmeas}{switch status}

\newcommand{\Statusmeass}{Switch status data}

\begin{document}

\title{A Convex Method of Generalized State Estimation using Circuit-theoretic Node-breaker Model}

\author{Shimiao~Li,~\IEEEmembership{Graduate Student Member,~IEEE,}
	Amritanshu~Pandey,~\IEEEmembership{Member,~IEEE,}
	and~Larry~Pileggi,~\IEEEmembership{Fellow,~IEEE}
\thanks{
Manuscript received 13 December 2022; revised 26 August 2023; accepted
12 November 2023. This work was supported in part by C3.ai Inc., and in part
by Microsoft Corporation. Paper no. TPWRS-01859-2022.\\

\copyright 2021 IEEE. Personal use of this material is permitted. Permission from IEEE must be obtained for all other uses, in any current or future media, including
reprinting/republishing this material for advertising or promotional purposes, creating new collective works, for resale or redistribution to servers or lists, or reuse of any copyrighted component of this work in other works.\\

S. L. and L. P. are with the Department of Electrical and Computer Engineering at Carnegie Mellon University, Pittsburgh, PA, 15213 USA (email:\{shimiaol,pileggi\}@andrew.cmu.edu).\\

A. P. is with the Department of Electrical Engineering at the University of Vermont, Burlington, VT. 05405 USA (email: amritanshu.pandey@uvm.edu).}
}

\maketitle

\begin{abstract}
	\input{000abstract}
\end{abstract}


\input{nomenclature}
\section{Introduction}
\label{sec:Introduction}
\input{010introduction}

\input{020background}

\section{Circuit-Theoretic Generalized State-Estimation}\label{sec:method}
\input{030method}

\section{Results}
\label{sec:Results}
\input{040experiment}

\input{050discussion}

\section{Conclusion}
\label{sec:Conclusion}
\input{060conclusion}


\bibliographystyle{IEEEtran}
\bibliography{ckt-GSE}

\input{appendix}

\input{biography}

\end{document}

%% file: 000abstract.tex
An accurate and up-to-date topology is critical for situational awareness of a power grid; however, wrong switch statuses due to physical damage, communication error, or cyber-attack, can often result in topology errors. To maintain situation awareness under the possible topology errors and bad data, this paper develops ckt-GSE, a circuit-theoretic generalized state estimation method using node-breaker (NB) model. Ckt-GSE is a convex and scalable model that jointly estimates AC state variables and network topology, with robustness against different data errors. The method first constructs an equivalent circuit representation of the AC power grid by developing and aggregating linear circuit models of SCADA meters, phasor measurement units(PMUs), and switching devices. Then based on this circuit, ckt-GSE defines a constrained optimization problem using weighted least absolute value (WLAV) objective to form a robust estimator. The problem is a Linear Programming (LP) problem whose solution includes accurate AC states and a sparse vector of noise terms to identify topology errors and bad data.This paper is the first to explore a circuit-theoretic approach for an AC-network constrained GSE algorithm that is: 1) applicable to the real-world data setting, 2) convex without relaxation, scalable with our circuit-based solver; and 3) robust with the ability to identify and reject different data errors.

%% file: nomenclature.tex
\section*{Nomenclature}
\addcontentsline{toc}{section}{Nomenclature}
\begin{IEEEdescription}[\IEEEusemathlabelsep\IEEEsetlabelwidth{$P_{rtu}, Q_{rtu}$}]
\item [$V_i^R, V_i^I$]  Real and imaginary voltage at bus $i$.
\item [$V_i$]  Complex voltage at bus $i$, $V_i=V_i^R+jV_i^I$.
\item [$x$]  AC State vector. $x=[V^R_1,V^I_1,...,V^R_N,V^I_N]$. 
\item [$P_i, Q_i$]  Real and reactive power injection at bus $i$.
\item [$P_{ij}, Q_{ij}$] Real and reactive power on line ($i$, $j$)
\item [$P_{rtu}, Q_{rtu}$]  $P,Q$ data given by RTU (SCADA meter).
\item [$|V|_{rtu}$] $|V|$ data given by RTU.
\item [$V_{pmu}, I_{pmu}$] Voltage and current phasor given by PMU,  $V_{pmu}=V_{pmu}^R+jV_{pmu}^I, I_{pmu}=I_{pmu}^R+jI_{pmu}^I$
\item [$I_{sw}$]  Current on switch (sw), $I_{sw} = I_{sw}^R+ I_{sw}^I$.
\item [$n_{pmu}, n_{rtu}$]  Slack variable to capture error on PMU, RTU.
\item [$n_{sw}$]  Slack variable to capture error of switch status, 
$n_{pmu/rtu/sw}=n_{pmu/rtu/sw}^R+jn_{pmu/rtu/sw}^I$.
\item [$G+jB$]  Admittance; $G$: conductance, $B$: susceptance.
\end{IEEEdescription}

All variables above are defined in per unit (p.u.).

%% file: 010introduction.tex
Grid operation relies on the situational awareness provided by AC state estimation (ACSE) algorithms. Existing ACSE algorithms, however, such as the weighted least square (WLS) method \cite{traditional-WLS-SE}\cite{sugar-se-L2}, are based on the bus-branch model of the grid that assumes an error-free input topology. As such, any error in the topology data will degrade the quality of the ACSE solution.

To obtain an accurate and up-to-date topology, a network topology processor (NTP) \cite{TE-NTP-book}\cite{TE-NTP-tracking} is used in the control rooms today. It transforms the input \statusmeass{} into a bus-branch model by identifying the network connectivity. However, NTP assumes that the input \statusmeass{} are all accurate and thus cannot account for topology errors due to communication delay, incorrect operator entry, physical damage, or cyber-attacks.

Later works have proposed advancements to overcome the challenges in NTP.  These methods are designed to identify anomalous \statusests{} and the resultant topology errors \cite{GSE-DC-substation}\cite{GSE-AC-nonlinear}\cite{TE-WLSE-BDI}\cite{TESE-GSE-PMUabur}\cite{convexTESE-SDP-weng}. One family of approaches, namely generalized topology processing \cite{GSE-DC-substation}, also known as the classical generalized state estimation (GSE), creates pseudo measurements for switching devices within a substation and runs local state estimation followed by hypothesis tests to detect wrong switch statuses within the small region. In another approach, a variant of ACSE bad-data detection (BDD) \cite{TE-WLSE-BDI} runs ACSE  for a set of (bus-branch model) topologies and then applies residual tests to determine the optimal topology based on the one with the smallest residual. More recently, advanced GSE methods \cite{TESE-GSE-PMUabur}\cite{convexTESE-SDP-weng}\cite{TESE-GSE} have been proposed that use a node breaker (NB) model to perform a joint estimation of AC states and topology for the entire AC power grid, allowing bad continuous data and topology error to be effectively identified and separated. However, existing methods have significant drawbacks that limit their efficacy for use in a real-world setting.
%
\begin{table*}[htbp]
\caption{Feature Comparison of State and Topology Estimation Methods}
\label{tab: features}
\setlength{\extrarowheight}{2pt}
\begin{tabular}{cc|ccc|cc|cccc|cc}
\hline
\multicolumn{2}{c|}{\multirow{3}{*}{Method}}                                                               & \multicolumn{3}{c|}{Data}                                                                                     & \multicolumn{2}{c|}{Estimation Target}                                   & \multicolumn{4}{c|}{Robust to Errors}                                                        & \multicolumn{2}{c}{Properties}                                                                                                 \\ \cline{3-13} 
\multicolumn{2}{c|}{}                                                                                      & \multicolumn{1}{c|}{\multirow{2}{*}{sw}} & \multicolumn{1}{c|}{\multirow{2}{*}{SCADA}} & \multirow{2}{*}{PMU} & \multicolumn{1}{c|}{\multirow{2}{*}{states}} & \multirow{2}{*}{topology} & \multicolumn{2}{c|}{bad data}                           & \multicolumn{2}{c|}{topology error} & \multicolumn{1}{c|}{\multirow{2}{*}{convex}}                             & \multirow{2}{*}{scalable}                           \\ \cline{8-11}
\multicolumn{2}{c|}{}                                                                                      & \multicolumn{1}{c|}{}                    & \multicolumn{1}{c|}{}                       &                      & \multicolumn{1}{c|}{}                        &                           & \multicolumn{1}{c|}{identify} & \multicolumn{1}{c|}{reject} & \multicolumn{1}{c|}{identify} & reject & \multicolumn{1}{c|}{}                                                    &                                                     \\ \hline
\multicolumn{1}{c|}{\multirow{2}{*}{TE}}  & NTP \cite{TE-NTP-book}\cite{TE-NTP-tracking}                                                            & \multicolumn{1}{c|}{\color{blue} \textbf{\cmark} }                   & \multicolumn{1}{c|}{\color{red} \xmark}                      & {\color{red} \xmark}                    & \multicolumn{1}{c|}{{\color{red} \xmark}}                       & {\color{blue} \textbf{\cmark}}                         & \multicolumn{1}{c|}{{\color{red} \xmark}}    & \multicolumn{1}{c|}{{\color{red} \xmark}}      & \multicolumn{1}{c|}{{\color{red} \xmark}}    & {\color{red} \xmark}      & 
\multicolumn{1}{c|}{ /}                                                   & /                                             \\
\cline{2-13} 
\multicolumn{1}{c|}{}                     & Traditional BDI \cite{TE-WLSE-BDI}                                               & \multicolumn{1}{c|}{{\color{red} \xmark}}                   & \multicolumn{1}{c|}{{\color{blue} \textbf{\cmark}}}                      & {\color{blue} \textbf{\cmark}}                    & \multicolumn{1}{c|}{{\color{blue} \textbf{\cmark}}}                       & {\color{blue} \textbf{\cmark}}                         & \multicolumn{1}{c|}{{\color{red} \xmark}}    & \multicolumn{1}{c|}{{\color{red} \xmark}}      & \multicolumn{1}{c|}{{\color{blue} \textbf{\cmark}}}    & {\color{red} \xmark}      & \multicolumn{1}{c|}{{\color{red} \xmark}}                                                   & {\color{red} \xmark}                                                   \\ \hline
\multicolumn{1}{c|}{\multirow{3}{*}{ACSE}}  & Traditional SE\&BDI \cite{traditional-WLS-SE}                                           & \multicolumn{1}{c|}{{\color{red} \xmark}}                   & \multicolumn{1}{c|}{{\color{blue} \textbf{\cmark}}}                      & {\color{blue} \textbf{\cmark}}                    & \multicolumn{1}{c|}{{\color{blue} \textbf{\cmark}}}                       & {\color{red} \xmark}                         & \multicolumn{1}{c|}{{\color{blue} \textbf{\cmark}}}    & \multicolumn{1}{c|}{{\color{red} \xmark}}      & \multicolumn{1}{c|}{{\color{red} \xmark}}    & {\color{red} \xmark}      & \multicolumn{1}{c|}{{\color{red} \xmark}}                                                   & {\color{red} \xmark}                                                   \\ \cline{2-13} 
\multicolumn{1}{c|}{}                     & Robust linear SE, pmu\cite{abur-robustSE-PMU}                                          & \multicolumn{1}{c|}{{\color{red} \xmark}}                   & \multicolumn{1}{c|}{{\color{red} \xmark}}                      & {\color{blue} \textbf{\cmark}}                    & \multicolumn{1}{c|}{{\color{blue} \textbf{\cmark}}}                       & {\color{red} \xmark}                         & \multicolumn{1}{c|}{{\color{blue} \textbf{\cmark}}}    & \multicolumn{1}{c|}{{\color{blue} \textbf{\cmark}}}      & \multicolumn{1}{c|}{{\color{red} \xmark}}    & {\color{red} \xmark}      & \multicolumn{1}{c|}{}                                                    &                                                     \\ \cline{2-13} 
\multicolumn{1}{c|}{}                     & Robust ckt-SE \cite{SUGAR_SE_WLAV}                                                 & \multicolumn{1}{c|}{{\color{red} \xmark}}                   & \multicolumn{1}{c|}{{\color{blue} \textbf{\cmark}}}                      & {\color{blue} \textbf{\cmark}}                    & \multicolumn{1}{c|}{{\color{blue} \textbf{\cmark}}}                       & {\color{red} \xmark}                         & \multicolumn{1}{c|}{{\color{blue} \textbf{\cmark}}}    & \multicolumn{1}{c|}{{\color{blue} \textbf{\cmark}}}      & \multicolumn{1}{c|}{{\color{red} \xmark}}    & {\color{red} \xmark}      & \multicolumn{1}{c|}{{\color{blue} \textbf{\cmark}}}                                                   & {\color{blue} \textbf{\cmark}}                                                   \\ \hline
\multicolumn{1}{c|}{\multirow{4}{*}{GSE}} & Classic GTP/GSE \cite{GSE-DC-substation}\cite{GSE-AC-nonlinear}                                           & \multicolumn{1}{c|}{{\color{blue} \textbf{\cmark}}}                   & \multicolumn{1}{c|}{{\color{blue} \textbf{\cmark}}}                      & {\color{blue} \textbf{\cmark}}                    & \multicolumn{1}{c|}{{\color{red} local}}                   & {\color{red} local}                     & \multicolumn{1}{c|}{{\color{blue} \textbf{\cmark}}}    & \multicolumn{1}{c|}{{\color{red} \xmark}}      & \multicolumn{1}{c|}{{\color{blue} \textbf{\cmark}}}    & {\color{red} \xmark}      & \multicolumn{1}{c|}{\begin{tabular}[c]{@{}c@{}}DC\cite{GSE-DC-substation}-{\color{blue} \textbf{\cmark}}\\ AC\cite{GSE-AC-nonlinear}-{\color{red} \xmark}\end{tabular}} & \begin{tabular}[c]{@{}c@{}}DC-{\color{blue} \textbf{\cmark}}\\ AC-{\color{red} \xmark}\end{tabular} \\ \cline{2-13} 
\multicolumn{1}{c|}{}                     & GSE-pmu \cite{TESE-GSE-PMUabur}                                                        & \multicolumn{1}{c|}{{\color{blue} \textbf{\cmark}}}                   & \multicolumn{1}{c|}{{\color{red} \xmark}}                      & {\color{blue} \textbf{\cmark}}                    & \multicolumn{1}{c|}{{\color{blue} \textbf{\cmark}}}                       & {\color{blue} \textbf{\cmark}}                         & \multicolumn{1}{c|}{{\color{blue} \textbf{\cmark}}}    & \multicolumn{1}{c|}{{\color{blue} \textbf{\cmark}}}      & \multicolumn{1}{c|}{{\color{blue} \textbf{\cmark}}}    & {\color{blue} \textbf{\cmark}}      & \multicolumn{1}{c|}{{\color{blue} \textbf{\cmark}}}                                                   & {\color{blue} \textbf{\cmark}}                                                   \\ \cline{2-13} 
\multicolumn{1}{c|}{}                     & GSE-SDP \cite{convexTESE-SDP-weng}                                                        & \multicolumn{1}{c|}{{\color{blue} \textbf{\cmark}}}                   & \multicolumn{1}{c|}{{\color{blue} \textbf{\cmark}}}                      & {\color{blue} \textbf{\cmark}}                    & \multicolumn{1}{c|}{{\color{blue} \textbf{\cmark}}}                       & {\color{blue} \textbf{\cmark}}                         & \multicolumn{1}{c|}{{\color{blue} \textbf{\cmark}}}    & \multicolumn{1}{c|}{{\color{blue} \textbf{\cmark}}}      & \multicolumn{1}{c|}{{\color{blue} \textbf{\cmark}}}    & {\color{blue} \textbf{\cmark}}      & \multicolumn{1}{c|}{{\color{blue} \textbf{\cmark}}}                                                   & {\color{red} \xmark}                                                   \\ \cline{2-13} 
\multicolumn{1}{c|}{}                     & \begin{tabular}[c]{@{}c@{}}{\color{blue} \textbf{ckt-GSE}}\\ {\color{blue} \textbf{(this paper)}}\end{tabular} & \multicolumn{1}{c|}{{\color{blue} \textbf{\cmark}}}                   & \multicolumn{1}{c|}{{\color{blue} \textbf{\cmark}}}                      & {\color{blue} \textbf{\cmark}}                    & \multicolumn{1}{c|}{{\color{blue} \textbf{\cmark}}}                       & {\color{blue} \textbf{\cmark}}                         & \multicolumn{1}{c|}{{\color{blue} \textbf{\cmark}}}    & \multicolumn{1}{c|}{{\color{blue} \textbf{\cmark}}}      & \multicolumn{1}{c|}{{\color{blue} \textbf{\cmark}}}    & {\color{blue} \textbf{\cmark}}      & \multicolumn{1}{c|}{{\color{blue} \textbf{\cmark}}}                                                   & {\color{blue} \textbf{\cmark}}                                                   \\ \hline
\end{tabular}
\footnotesize{*sw: circuit breaker/switch status data.\\
* 'identify' means detect and localize the error, 'reject' means automatically removing the impact of erroneous data so that it does not affect the results.\\
* Table \ref{tab: theoretical comparison} compares ckt-GSE with GSE-pmu and GSE-SDP in more detail.}
\end{table*}

The classical GSE algorithm \cite{GSE-DC-substation} was demonstrated on a small substation network using a linear DC grid model. When extended to nonlinear AC network models of power grids\cite{GSE-AC-nonlinear}, the problem becomes non-convex and NP-hard to solve. The approach in \cite{TE-WLSE-BDI} that adopts WLS estimation and identifies anomalous topology from residual-based hypothesis tests is also non-convex and can fail when multiple switch statuses are erroneous concurrently. More advanced works \cite{TESE-GSE-PMUabur}\cite{GSE-AC-nonlinear}\cite{convexTESE-SDP-weng}, which proposed NB-model based AC network-constrained GSE formulations have challenges as well. Among these works, \cite{TESE-GSE-PMUabur} assumes full observability of the network model using PMU data alone; however, this assumption is unrealistic in real-world grids where traditional SCADA RTU meters are still dominant. In separate work, \cite{convexTESE-SDP-weng} uses semidefinite programming (SDP) relaxation to obtain a convex GSE model, but SDP does not scale well to large-scale systems \cite{large-scale-SDP}. {\color{black} While most existing models are unconstrained, some other works\cite{constrained-GSE1}\cite{constrained-GSE2} are built on constrained optimization problems with zero-injection buses included in equality constraints; however, most works \cite{constrained-GSE1}\cite{constrained-GSE2}  do not guarantee convexity and  are not robust estimators.}

To motivate our method, we summarize the features and drawbacks of current  techniques in Table \ref{tab: features}. We find that challenges exist in existing models in terms of i) input data applicability, ii) robustness to data errors, and iii) convergence guarantee and scalability.

To address these challenges, this paper proposes \textbf{ckt-GSE}, a novel AC-network constrained generalized state estimation (GSE) algorithm with a circuit-theoretic foundation. For comparison to other methods, the features of ckt-GSE are shown in the last row of Table \ref{tab: features}. 
{\color{black}  These features also motivate an informal task definition of ckt-GSE:

\begin{definition}[\textbf{ckt-GSE task}] Given a snapshot of (redundant) data that guarantees system observability, ckt-GSE maps measurement data into an aggregated circuit model to formulate a constrained optimization problem and obtain a joint estimation of power grid AC states and topology, with nice properties of convexity 
(numerical stability), scalability, and robustness against multiple data errors. The collected data include 1) \statusmeass{} that are discrete measurements indicating on/off, and 2) \contmeass{} of voltage, current, and power measured by PMUs and SCADA (RTUs). Both types of data can include random noise and errors.
\label{def: ckt-GSE}
\end{definition}

As stated in the definition, ckt-GSE relies highly on the circuit-theoretic modeling of grid data. In \cite{SUGAR_SE_WLAV}, we developed linear circuit models for RTUs and PMUs to construct the power grid’s AC bus-branch (BB) model. However, the estimators on BB model does not include switching devices and thus cannot account for topology errors (i.e., wrong switch statuses).
Here to include topology estimation, we extend the previous work from BB model to node-breaker model, by developing the linear circuit models for open and closed switches to construct a node-breaker (NB) model of the grid. }All these models are updated upon the arrival of new grid data. Then with these linear circuit models of switches, PMUs and RTUs, ckt-GSE constructs an aggregated linear equivalent circuit to represent the up-to-date steady-state operating point of the entire grid in node-breaker (NB) settings. Kirchhoff’s circuit laws are then applied to develop an estimation approach using the circuit model of the grid.

We designed ckt-GSE as a robust estimator with convergence guarantees and scalability. Specifically, if measurements are ideal and error-free, the constructed aggregated circuit for the grid’s NB model will satisfy Kirchhoff’s Current Laws (KCL) at all nodes. However, real-world measurements can be erroneous. As such, data errors in the NB model will result in an illegal circuit for which KCL constraints will not be satisfied. To incorporate data error into the grid model, we introduce \textit{slack} injection current sources to all measurement models in the aggregated circuit to represent and capture data errors. The slack variables ensure that KCL constraints at all nodes are satisfied even when an error is present. They play the role of compensation to allow an automatic correction of data errors in the models. Then, we obtain a joint estimate of grid states and switch statuses by minimizing the weighted least absolute value (WLAV) of slack injection sources. The resulting ckt-GSE is a  \textbf{Linear Programming (LP)}  problem characterized by WLAV objective and linear network constraints. The solution includes a sparse vector of slack current values to identify and localize suspicious  \statusmeass{} and bad \contmeass. Meanwhile, the robust estimator automatically rejects erroneous data, ensuring that the estimates of system states are reliable.

To the best of our knowledge, this paper is the first to develop a circuit-theoretic approach for AC GSE that considers realistic data collection and is also convex, scalable, and robust to different data errors. We show the efficacy of our approach on node-breaker test cases of different sizes.

%% file: 020background.tex
\section{Related Work}\label{sec:background}
\subsection{Node-Breaker (NB) model}
\begin{figure}[h]
	\centering
	\includegraphics[width=1.0\linewidth]{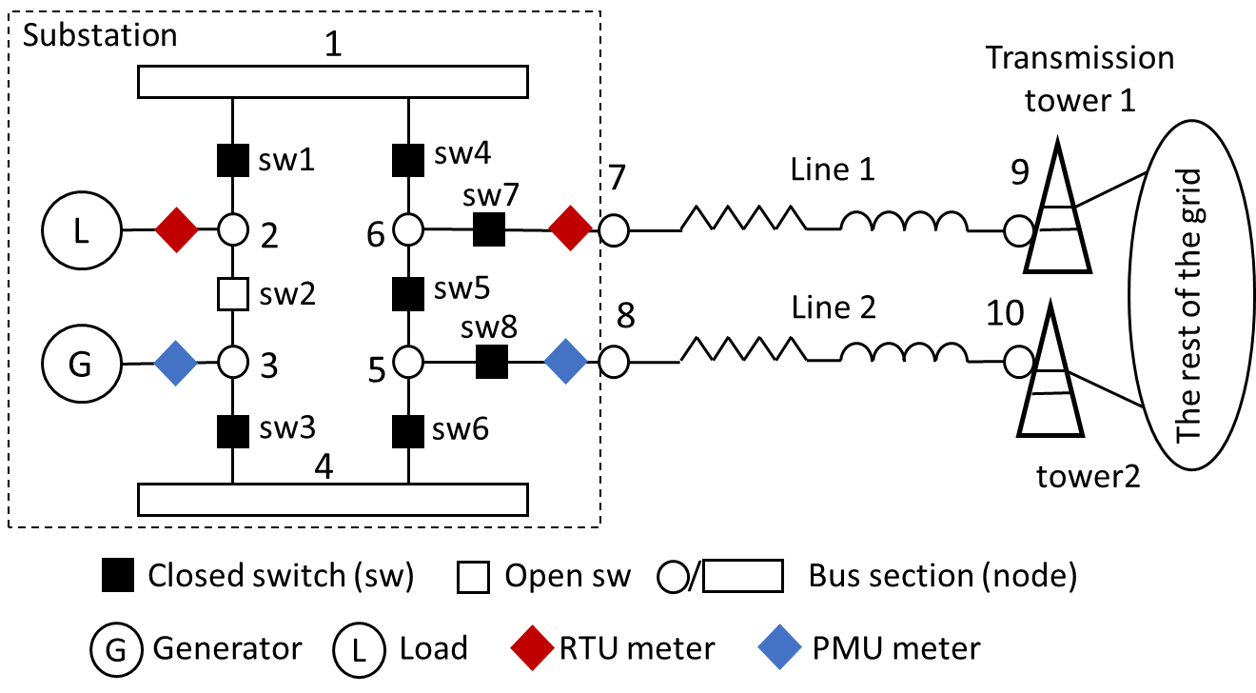}
	\caption[]{Toy example: the node breaker (NB) model of a grid sub-network around a substation.
}
	\label{fig:NB model}
\end{figure}

Unlike the bus-branch (BB) model which is a simplified representation of the power grid, the node-breaker (NB) model, also called the bus section/circuit breaker model, gives a complete physical description of the power grid to enable direct consideration of each grid component. Specifically, it contains bus sections (i.e., nodes) at different voltage levels, circuit breakers and switches,  locations of metering devices, as well as other equipment that are included in the BB model, such as transmission lines, transformers, shunts, generators and loads, etc. Fig. \ref{fig:NB model} shows a node-breaker model of a grid sub-network in the vicinity of a substation.

\subsection{Network Topology Processor (NTP)}

While the NB model represents the actual power system more comprehensively, it includes many inactive and active switch components which increase the size of the network and subsequently the run-time for analysis. Therefore, the network topology processor (NTP) unit converts the NB model into a bus-branch format, which is used by today's AC state estimator and other subsequent units. NTP removes inactive switches and shorts active switches to construct a model with only the active buses, lines and transformers. The standard algorithm for NTP\cite{TE-NTP-book} is as follows:
\begin{enumerate}
    \item \textit{Raw data processing} converts raw data into normalized units and performs simple checks to verify operating limits, rate of change of operating variables, and other data consistency checks (e.g., confirm zero flows on open switches and zero voltage across closed switches).
    \item \textit{Bus section processing} recursively merges any two nodes connected by a closed switch into one bus.
    \item \textit{Connectivity analysis} identifies active network topology from the \statusmeass{} and reassigns locations of metering devices on the bus-branch model.
\end{enumerate}
As only minor or even no topology changes occur most of the time between two subsequent NTP processes, a real-world NTP increases its efficiency by operating in tracking mode\cite{TE-NTP-tracking} where only the sub-networks where topology changes occur are processed and updated recursively. 

When processing on the NB model, NTP assumes all its input status data are correct. However, \statusmeas{} in a practical grid setting can be corrupted due to telemetry error, operator entry error, physical damage (e.g., a line is tripped but the disconnection is not reflected on the circuit breaker status) or even a cyber-attack. In case of such incorrect \statusmeas{}, NTP can falsely merge or split buses and output an erroneous grid topology. 

\subsection{Generalized state estimation (GSE) on a Node-Breaker Model}
As discussed, NTPs do not account for possible errors in \statusmeass{} when constructing the grid topology. Therefore, any erroneous topology from NTP will be fed directly into traditional AC-state estimation (ACSE). While bad-data detection algorithms within ACSE can detect random bad continuous data, they are not designed to detect topology errors. Thus, any topology error in today's control room setting can negatively impact the real-time operation.

As a result, a more robust state estimation called the generalized state-estimation approach has been introduced \cite{GSE-DC-substation}, \cite{GSE-AC-nonlinear}. Originally, \cite{GSE-DC-substation} proposed a DC network-constrained version of GSE (DC-GSE), which runs on an NB model. The mechanism of the algorithm is as follows:
\begin{enumerate}
    \item \textit{Modeling of switches}: For any switch $(sw)$ that connects node $i,j$ in the NB model, \cite{GSE-DC-substation} creates a pseudo measurement of zero power flow ($P_{ij}=0$) if it is open, and zero angle difference ($\theta_{ij}=0$) if it is closed. Other elements are modeled similarly to the bus-branch model.
    \item \textit{Estimation}: These pseudo measurements for discrete states, along with the \contmeass{}, are then used to run DC-GSE on the network.
    \item \textit{Bad-data Detection}: A hypothesis test is performed on the residual of each switch and continuous measurement to check if any data is wrong. 
\end{enumerate}
 While the use of a DC model provides the desirable properties of linearity and problem convexity, it does not have the expressiveness or fidelity to represent the AC system accurately. Therefore, \cite{GSE-AC-nonlinear} extended the DC-GSE to AC-constrained GSE. But it has challenges as well. Due to nonlinear branch flow equations, AC-GSE results in a non-convex formulation with significant drawbacks in performance \cite{GSE-AC-nonlinear}. A recent work \cite{TESE-GSE-PMUabur} formulated a convex GSE problem with AC constraints under rectangular coordinates. However, this method only uses PMU measurements and is not generalizable to measurements ($P_{rtu},Q_{rtu},|V|_{rtu}$) collected by RTUs, which are the prevalent meters in today's grid. Work in \cite{convexTESE-SDP-weng} formulated a robust GSE model with both SCADA RTU and PMU considered in the input data, however, the use of semidefinite programming (SDP) relaxation to convexify the problem results in a lack of scalability and efficiency on large-scale networks. 

{\color{black} Further, in terms of the type of optimization problem that GSE solves, most of the methods discussed above, like \cite{TESE-GSE-PMUabur}\cite{convexTESE-SDP-weng}\cite{TESE-GSE}, are based on solving unconstrained optimization problems which minimize the measurement error. However, there also exist some works, like \cite{constrained-GSE1}\cite{constrained-GSE2}, that developed constrained GSE (CGSE) models which are constrained optimization problems that include zero-injection buses, switch statuses (or circuit breaker flows) and some measurements  in the equality constraint set. However, these works still suffer from nonlinearity when considering SCADA (RTU) data, and many works\cite{constrained-GSE1}\cite{constrained-GSE2} adopted a weighted least square objective and thus are not robust estimators.}
 

{\color{black}
\subsection{Learning-based topology estimators}

Beyond these physical model-based methods, recent years have also witnessed a large number of data-driven or learning-based approaches to reconstruct the structure of power grids. Most of them are designed for the distribution networks, specifically to handle the radial (tree-like) networks, and the sparse installation of monitoring devices which makes direct observation and estimation challenging.
Existing methods include, but are not limited to, time-series pattern recognition methods \cite{TE-patternrecg-4dsit}, probabilistic graphical models (\cite{TE-UGM-4dist} deployed an undirected graphical model, \cite{TE-BayesianNet-4dist} deployed a Bayesian Network, .etc), tree-based methods\cite{TE-tree-4dist}, .etc.  
Many of these methods require high-precision and high-frequency synchronous measurements as input. 
Also, learning-based approaches which require modeling training will unavoidably suffer from generalization issues, raising the fear of giving inaccurate predictions on unseen data. And limited by the training complexity, many works are only demonstrated on small-sized networks and fail to scale well. 
For reliable performance on any unseen power grid, this work, as stated in Definition. \ref{def: ckt-GSE}, still focuses on physical model-based estimation from a snapshot of measurements which guarantee 100\% observability (mainly on transmission networks). 
Thus those learning based approaches are out of the scope of this work. 
}

%% file: 030method.tex

To address the drawbacks of today's GSE algorithms, we develop \textbf{ckt-GSE}, a novel circuit-theoretic approach for the AC-network constrained GSE problem. The foundation of the 
circuit-theoretic framework lies in constructing an aggregated equivalent circuit of the power grid whose elements are characterized by their I-V relationship \cite{sugar-powerflow-amrit}. The approach can map all grid components into corresponding equivalent circuits without loss of generality, including measurement data from grid sensors. An optimization problem can be defined thereafter on the aggregated circuit to perform certain target functionality, like optimal power flow analysis \cite{opf} and state estimation\cite{sugar-se-L2}\cite{SUGAR_SE_WLAV}. \textit{Optimization and analysis on the circuit representation are equivalent to those on the original AC system because the developed circuit models capture AC grid physics without relaxation.} Next, to develop the ckt-GSE framework, we describe the construction of aggregated circuit with measurement data and an optimization algorithm that actuates on it.

\subsection{Equivalent circuit modeling of \statusmeass{} and \contmeass}\label{sec: models}

For ckt-GSE, we consider realistic grid settings with both \contmeass{} (from RTUs and PMUs) and discrete \statusmeass{} (from circuit-breakers). We build a linear circuit model for every measured component that captures its physics at the current operating point. Such a measurement-based model is updated recursively based on the latest data so that the \textit{up-to-date system physics} is captured accurately. 

\subsubsection{Linear models for PMU and RTU}

\begin{figure}[h]
     \centering
     \begin{subfigure}[b]{\linewidth}
         \centering         \includegraphics[width=0.8\linewidth]{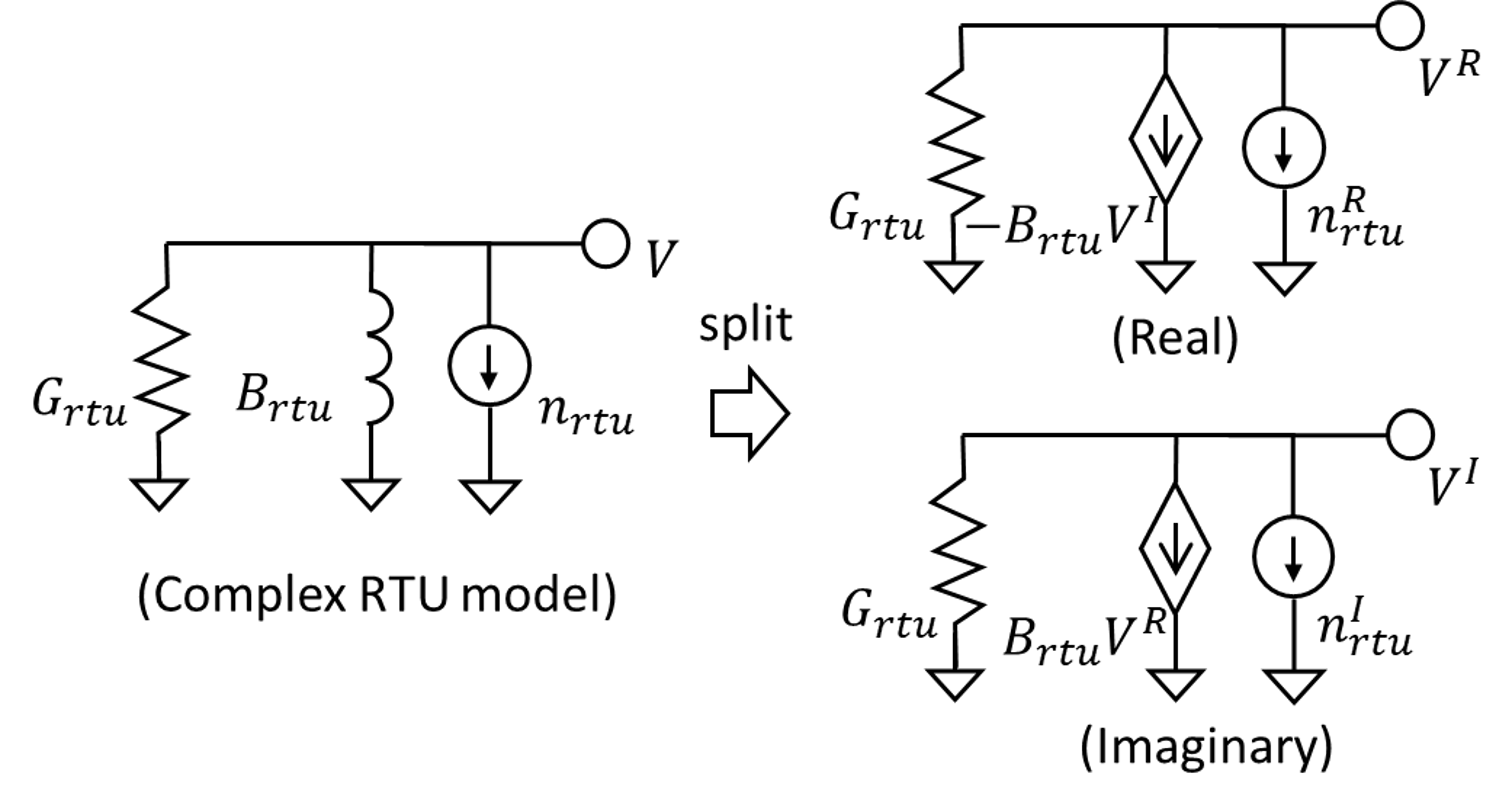}
         \caption{Linear RTU model: measurements are mapped to sensitivities.}
         \label{fig:RTU}
     \end{subfigure}
     \hfill
     \begin{subfigure}[b]{\linewidth}
         \centering         \includegraphics[width=1\linewidth]{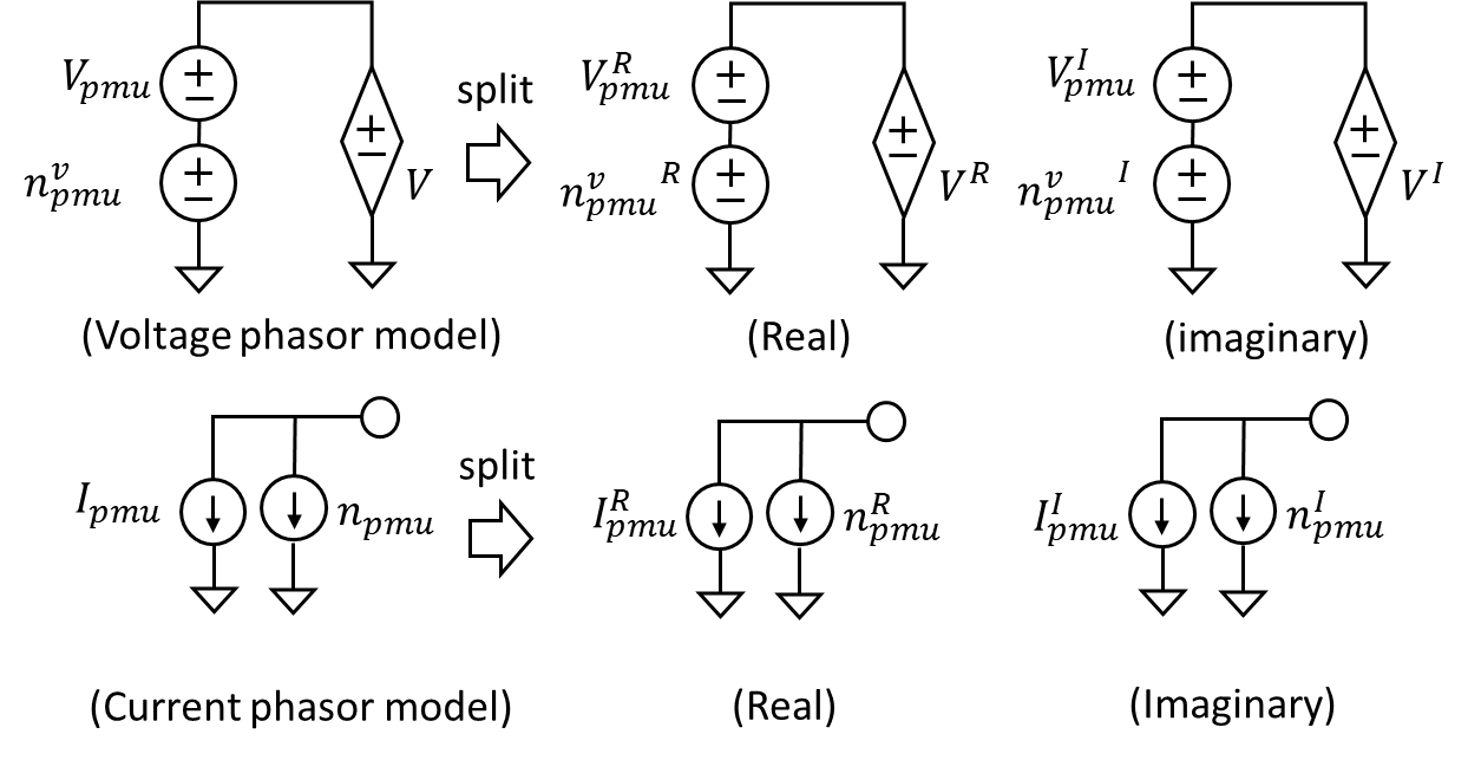}
         \caption{Linear PMU model: measurements have a linear nature under ECF.}
         \label{fig:PMU}
     \end{subfigure}
        \caption{RTU and PMU injection models.}
        \label{fig:models}
\end{figure}

Our prior works \cite{sugar-se-L2},\cite{SUGAR_SE_WLAV}\cite{sugar-SE-Hug}\cite{sugar-SE-Hug2} have developed linear models for PMU and RTU meters for AC-state estimation on the bus-branch model. Figure \ref{fig:RTU} shows the circuit model for RTU measurement devices, which are installed at a bus to measure bus voltages and power injections. In this model, the original measurements (bus voltage magnitude $|V|_{rtu}$, and power injections $P_{rtu},Q_{rtu}$) are mapped into sensitivities, $G_{rtu}$ and $B_{rtu}$:
\begin{equation}
    G_{rtu}=\frac{P_{rtu}}{|V|_{rtu}^2}; B_{rtu}=-\frac{Q_{rtu}}{|V|_{rtu}^2}
\end{equation} 

\noindent With these sensitivities, we can replace the measured sub-circuit with an RTU linear circuit model in Fig. \ref{fig:RTU}. {\color{black}Appendix \ref{apdx: RTU model} gives a more detailed expression of how circuit modeling serves to transform nonlinear measurements in a physically meaningful way for convex and linear constraint formulation.}

PMU measurements include current phasor $(I_{pmu}=I_{pmu}^{R}+jI_{pmu}^{I})$ and voltage phasor $(V_{pmu}=V_{pmu}^{R}+jV_{pmu}^{I})$  injection measurements. 
These measurements are intrinsically linear in the rectangular coordinate I-V framework, and the linear circuit model is characterized by independent current and voltage sources (see Figure \ref{fig:PMU}). Models for line flow measurements from RTUs and PMUs have been created similarly. To account for measurement errors, all models include slack variables $n^R, n^I$ to capture the measurement error. These variables capture both the real and the imaginary part of the noise/error, i.e, $n=n^R+j n^I$. See \cite{sugar-se-L2},\cite{SUGAR_SE_WLAV} for more details.

These models are directly applicable to the NB grid model as well. Fig. \ref{fig:NB circuit model} shows a schematic where any component measured by these devices (in Fig. \ref{fig:NB model}) is replaced by a linear circuit counterpart.

Note that one might question whether the linear $G_{rtu}$ and $B_{rtu}$ parameters used to model the RTU measurements at load buses are relaxations of the actual $P_{rtu}$ and $Q_{rtu}$ measurements. However, from an equivalent circuit perspective, we argue that our circuit modeling is \textbf{not a relaxation}. From a single measurement, one cannot gauge whether the measured load is constant power, constant impedance or ZIP load. So \textbf{all characterizations of loads are valid and equivalent representations of the current steady-state operating point}.


\subsubsection{Linear models for Switching Devices}
To perform GSE on an AC-network constrained NB model, this paper introduces two new models: i) open switch and ii) closed switch. The model considers the possibility of \textit{wrong} switch status (i.e., open switch reported as close and vice versa) and includes noise terms to estimate the correct switch status.

\begin{figure}[h]
	\centering
	\includegraphics[width=0.8\linewidth]{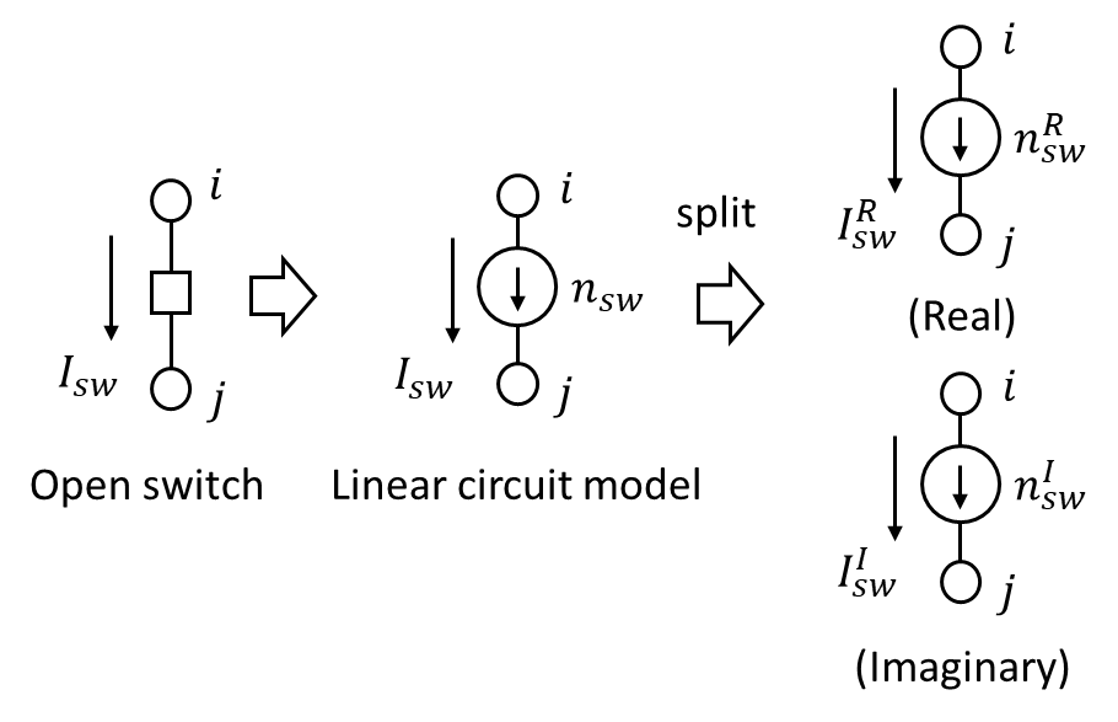}
	\caption[]{Open switch model: $n_{sw}$ close to zero if the status is correct; $n_{sw}$ compensates the current flow on the branch if the status is wrong.}
	\label{fig:open sw model}
\end{figure}

An open switch is simply modeled by an open circuit and as such no current can flow through it, i.e., the total current flow $I_{sw}=I_{sw}^R+jI_{sw}^I=0$. However, to account for a possibly wrong status, we add a slack current source in parallel, i.e. a noise term $n_{sw}=n_{sw}^R+jn_{sw}^I$, to compensate for the current that would otherwise flow through the switch in case it was actually closed. Fig. \ref{fig:open sw model} and (\ref{eq:open sw}) show the model where $R/I$ denote the real/imaginary parts.
\begin{equation}\label{eq:open sw}
    I_{sw}^R=n_{sw}^R; I_{sw}^I=n_{sw}^I
\end{equation}

\begin{figure}[h]
	\centering
	\includegraphics[width=1\linewidth]{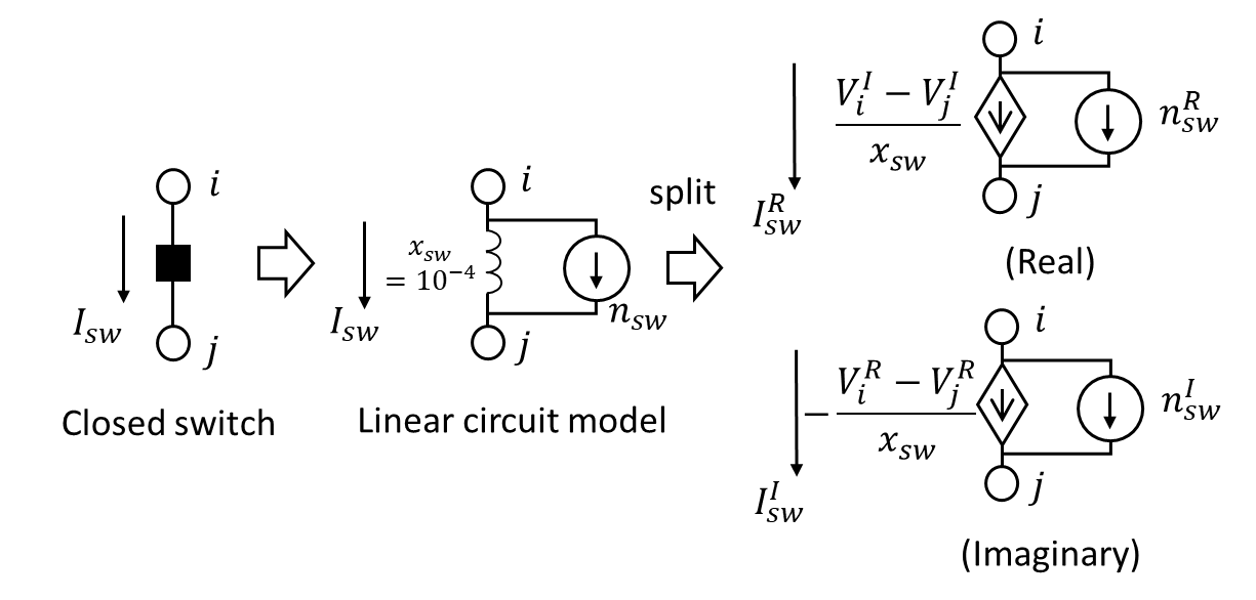}
	\caption[]{Closed switch model: $n_{sw}$ close to zero if the status is correct; $n_{sw}$ will offset the current flow on the branch if the status is wrong. $x_{sw}$ is in p.u.}
	\label{fig:closed sw model}
\end{figure}

A closed switch is modeled as a low impedance branch (reactance $x_{sw} \approx 0.0001$ p.u.), since a closed switch is ideally a short circuit with zero voltage drop across it. Similarly, to account for possibly wrong status, we add a slack current source (i.e., a 'noise' term) in parallel. In case the closed switch is actually open, the noise term $n_{sw}$ will provide sufficient current to nullify the current flowing through the closed switch model, such that the total current flow between the from and to node of the switch is zero, effectively representing an open switch. 
Fig. \ref{fig:closed sw model} shows the closed switch model, which is mathematically expressed in (\ref{eq:closed sw}).

\begin{equation}\label{eq:closed sw}
    I_{sw}^R=\frac{1}{x_{sw}}(V_i^I-V_j^I) + n_{sw}^R;  I_{sw}^I=-\frac{1}{x_{sw}}(V_i^R-V_j^R) + n_{sw}^I
\end{equation}

\subsection{Equivalent circuit modeling of other grid devices}\label{sec: physical_models}

To construct the overall NB model of the grid for ckt-GSE, we aggregate both the circuit models for the measurement devices and the physical devices such as transformers, lines, shunts, etc. All of these models are linear and their derivations and construction are covered in detail in \cite{sugar-powerflow-amrit}. Any nonlinear physical model (i.e., load or generation), for the purposes of ckt-GSE, is replaced by its equivalent linear circuit model from the measurement devices described in Section \ref{sec: models} following the substitution theorem in circuit-theory \cite{larry_book}.

\subsection{WLAV formulation of AC-constrained GSE problem}\label{sec:GSE formulation}


\begin{figure*}[h]
	\centering
	\includegraphics[width=0.8\linewidth]{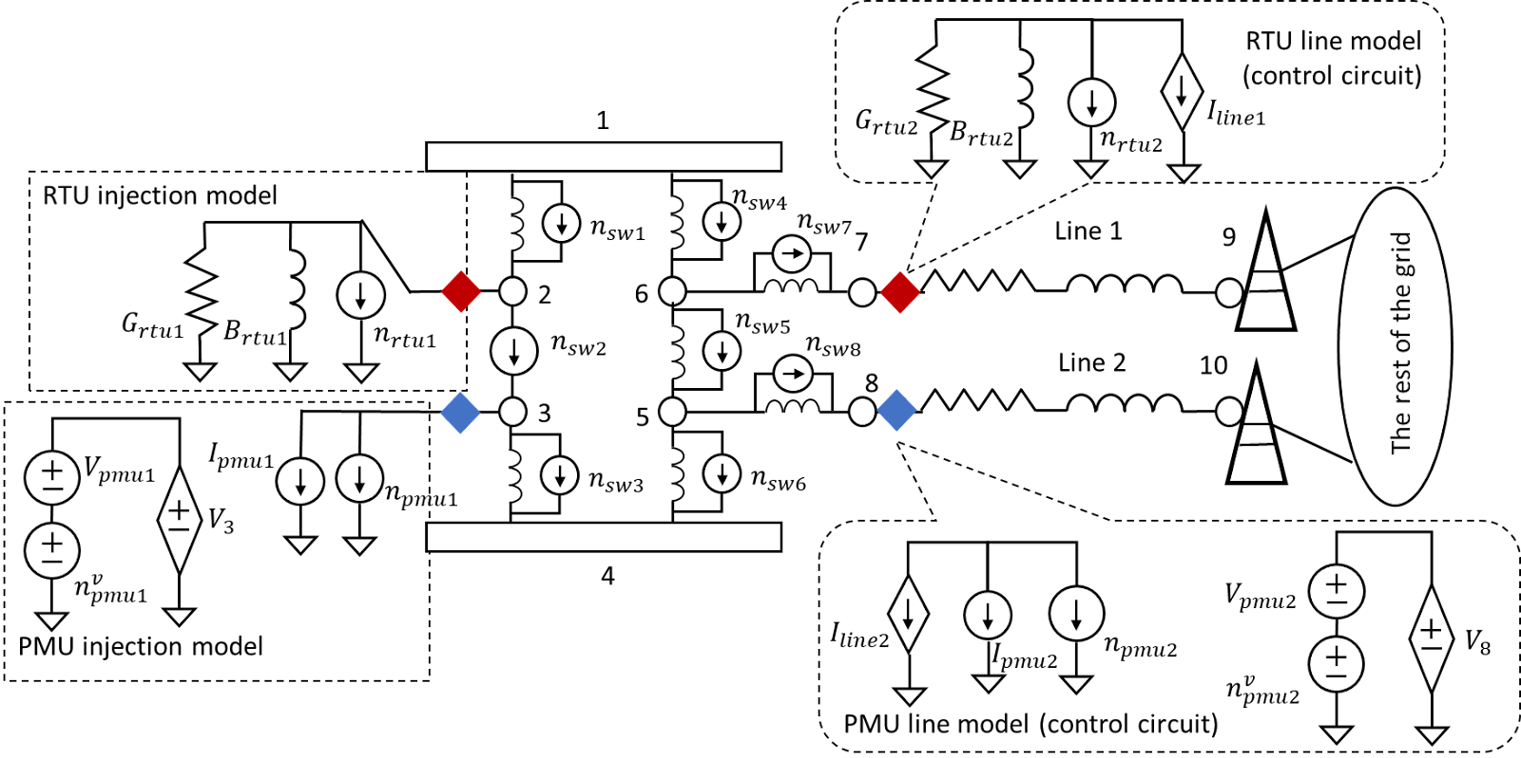}
	\caption[]{The NB example in Fig \ref{fig:NB model} is converted to a linear circuit after replacing grid components with their circuit models. To save space, this figure shows the complex models. It can be further split into real and imaginary parts by using the split circuit models in Section \ref{sec: models} and \ref{sec: physical_models} 
}
	\label{fig:NB circuit model}
\end{figure*}

With the circuit models described in Section \ref{sec: models} and Section \ref{sec: physical_models},  Fig. \ref{fig:NB circuit model} shows the equivalent circuit representation of an NB model in Fig. \ref{fig:NB model}, where we replace all switches and measured components with the established linear circuit models. Following the substitution theorem, the entire system is mapped to a linear circuit whose network constraints represented by Kirchhoff's current law (KCL) on all nodes, are a set of affine constraints.

The resulting aggregated circuit consisting of the main circuit and a set of control circuits captures the information from measurement data from an equivalent circuit-theoretic viewpoint. The main circuit captures the \textit{non-redundant set} of measurements including the AC network constraints at zero-injection nodes, whereas the set of control circuits captures the information from remaining redundant measurement data.

While a grid state can be obtained  by solving the network equations of the constructed linear circuit, infinitely many solutions exist as the introduction of slack variables $n^R, n^I$ results in an under-determined system of equations. Therefore, we next formulate an optimization problem to estimate a unique grid state, which provides a good estimate of grid states. Akin to conventional GSE methods \cite{GSE-DC-substation}\cite{GSE-AC-nonlinear}, we can formulate the optimization with a weighted least-square (WLS) algorithm, minimizing the L2-norm of the noise terms. However, in presence of bad data, the WLS formulation is not robust: it does not produce accurate estimates and requires post-processing to isolate suspicious measurements, followed by iteratively re-running the algorithm to obtain reliable estimates. Hence, to enhance the intrinsic robustness of ckt-GSE solution against wrong status and bad (continuous) data, we  choose to minimize the L1-norm of the noise terms subject to AC-network constraints at all nodes. This is equivalent to solving the weighted least absolute value (WLAV) estimation problem while satisfying the physics of the NB model:
\begin{subequations} \label{eq:GSE}
\begin{equation}
    \min_{x,n}\sum_k w_k|n_{swk}|+\sum_i\alpha_i|n_{rtui}|\notag\\
    + \sum_j\beta_{j}|n_{pmuj}| 
    + \gamma_{j}|n_{pmuj}^{v}|
    \label{obj}
\end{equation}
$$
    {\text{s.t. (linear) KCL equations at all nodes: } h(x,n)=0}
$$
{\color{black}
Taking the NB model in Fig \ref{fig:NB circuit model} as an example, the KCL equations are written as (\ref{eq: kcl example start})-(\ref{eq: kcl example end}):\\

\noindent Node 1: (connects switches)
\begin{equation}
\begin{split}
    I_{sw1}^R + I_{sw4}^R = 0, 
    I_{sw1}^I + I_{sw4}^I = 0
    \label{eq: kcl example start}
\end{split}
\end{equation}
Node 2: (connects RTU and switches)
\begin{equation}
\begin{split}
    G_{rtu1} V_2^R-B_{rtu1}V_2^I+n_{rtu1}^R + I_{sw2}^R - I_{sw1}^R = 0\\
    G_{rtu1}V_2^I+B_{rtu1}V_2^R+n_{rtu1}^I +  I_{sw2}^I - I_{sw1}^I = 0
\end{split}
\end{equation}
Node 3: (connects PMU and switches)
\begin{equation}
\begin{split}
    I_{pmu1}^R + n_{pmu1}^R - I_{sw2}^R+ I_{sw3}^R =0\\
     I_{pmu1}^I + n_{pmu1}^I - I_{sw2}^I+ I_{sw3}^I =0
\end{split}
\end{equation}
PMU voltage at node 3: (control circuit)
\begin{equation}
\begin{split}
    V_{pmu1}^R + n_{pmu1}^R - V_3^R = 0\\
    V_{pmu1}^I + n_{pmu1}^I - V_3^I = 0
\end{split}
\end{equation}
Node 4: (connects switches)
\begin{equation}
\begin{split}
I_{sw3}^R+I_{sw6}^R = 0, I_{sw3}^I+I_{sw6}^I=0
\end{split}
\end{equation}
Node 5: (connects switches)
\begin{equation}
\begin{split}
I_{sw6}^R+I_{sw8}^R-I_{sw5}^R = 0, I_{sw3}^I+I_{sw6}^I-I_{sw5}^I=0
\end{split}
\end{equation}
Node 6: (connects switches)
\begin{equation}
\begin{split}
I_{sw5}^R+I_{sw7}^R-I_{sw4}^R = 0, I_{sw5}^I+I_{sw7}^I-I_{sw4}^I=0
\end{split}
\end{equation}
Node 7: (connects switch and line)
\begin{equation}
\begin{split}
    G_{line1}(V_7^R-V_9^R) - B_{line1}(V_7^I-V_9^I)-I_{sw7}^R = 0\\
    G_{line1}(V_7^I-V_9^I) + B_{line1}(V_7^R-V_9^R) - I_{sw7}^I = 0
\end{split}
\end{equation}    
RTU-measured line (7,9): (control circuit)
\begin{equation}
\begin{split}
       G_{line1}(V_7^R-V_9^R) - B_{line1}(V_7^I-V_9^I) + G_{rtu2}V_7^R-\\B_{rtu2}V_7^I+n_{rtu2}^R = 0 \\
    G_{line1}(V_7^I-V_9^I) + B_{line1}(V_7^R-V_9^R) + G_{rtu2}V_7^I+\\B_{rtu2}V_7^R+n_{rtu2}^I =0\\
\end{split}
\end{equation}
Node 8: (connects switch and line)
\begin{equation}
\begin{split}
G_{line2}(V_8^R-V_{10}^R) - B_{line2}(V_8^I-V_{10}^I) - I_{sw8}^R = 0\\
G_{line2}(V_8^I-V_{10}^I) + B_{line2}(V_8^R-V_{10}^R) - I_{sw8}^I = 0\\
\end{split}
\end{equation}
PMU-measured line (8,10): (control circuit)
\begin{equation}
    G_{line2}(V_8^R-V_{10}^R) - B_{line2}(V_8^I-V_{10}^I) + I_{pmu2}^R + n_{pmu2}^R = 0
\end{equation}
\begin{equation}
    G_{line2}(V_8^I-V_{10}^I) + B_{line2}(V_8^R-V_{10}^R) + I_{pmu2}^I + n_{pmu2}^I = 0
\end{equation}
\begin{equation}
    V_{pmu2}^R + n_{pmu2}^R - V_8^R = 0
\end{equation}
\begin{equation}
    V_{pmu2}^I + n_{pmu2}^I - V_8^I = 0
\end{equation}
Open switches $sw_2$:  (control circuit):
\begin{equation}
    I_{sw2}^R = n_{sw2}^R, I_{sw2}^I = n_{sw2}^I
\end{equation}
Closed switches $sw_1,sw_3,sw_4,sw_5, sw_6, sw_7, sw_8:$ (control circuit for $swk=(i,j)$)
\begin{equation}
\begin{split}
    I_{swk}^R = \frac{1}{x_{sw}}(V_i^I-V_j^I) + n_{swk}^R,\\ 
    I_{swk}^I = -\frac{1}{x_{sw}}(V_i^R-V_j^R) + n_{swk}^I
\end{split}
\end{equation}
Other nodes in the system:
\begin{equation}
    ... 
    \label{eq: kcl example end}
\end{equation}
}
\end{subequations}

\noindent where the state vector $x=[V^R_1,V^I_1,...,V^R_N,V^I_N]$ contains real and imaginary bus voltages. $n_{rtu}, n_{pmu},$ and $n_{sw}$ represent the noise/error terms for RTUs, PMUs, and switches, respectively. Also,  ${w,\alpha,\beta,\gamma}$ are weights on each measurement model to represent a level of uncertainty, the selection of which will be discussed in Section \ref{sec: sensitivity}.

The use of the WLAV objective is inspired by the assumption that the data errors are sparsely distributed amongst the total measurement set since anomalies are rare in reality. 
As it minimizes the L1-norm objective, the WLAV estimator enforces a sparse vector of 'noise terms' that matches the sparse population of measurement errors. Large non-zero values only appear on locations with bad continuous data and wrong switch statuses, whereas the solution fits other high-quality measurements, providing robust estimates. 

Mathematically, the formulation in \eqref{eq:GSE} is a \textit{linear programming (LP)} problem, which is guaranteed to converge to a global optimum under the hold of certain conditions. The practical challenge stems from the non-differentiable L1 terms in the objective. To efficiently deal with the problem-solving, we first converted the objective function to a differential form:
\begin{subequations}
\label{prob: differentiable ckt-GSE}
\begin{equation}
\min_{x,n,t} c^Tt
\end{equation}
$$ s.t.\text{ network equations as in } (\ref{eq:GSE}) $$
\begin{equation}
    |n|\preceq t
    \label{con neq}
\end{equation}
\end{subequations}
\noindent with $c=[w,\alpha, \beta, \gamma]$, {\color{black} and the $t$ variable physically corresponds to the upper bound of the slack sources $n$.}

Then we adopt a circuit-theoretic LP solver by augmenting the standard primal-dual interior point (PDIP) algorithm with circuit-theoretic heuristics to speed up convergence. 
Specifically, the PDIP method solves the differentiable problem in (\ref{prob: differentiable ckt-GSE}) by iteratively solving the nonlinear perturbed KKT conditions {\color{black}  as follows:

\begin{subequations}
\noindent \text{Primal feasibility:}
\begin{align}
     Yx+Bn=J \text{ (linear KCL eqs)}\\
    |n|\preceq t
\end{align}
Complementary slackness:
\begin{align}
    \overline{\mu} (n-t) = -\epsilon\\
    \underline{\mu} (-n-t) = -\epsilon
\end{align}
Dual feasibility:
\begin{align}
   \mu \succeq 0, \mu = [\overline{\mu},\underline{\mu}]
\end{align}
\noindent Stationarity:
\begin{align}
    Y^T\lambda = 0\\
    \overline{\mu}-\underline{\mu} + B^T\lambda = 0\\
    \overline{\mu}+\underline{\mu} = c
\end{align}
\end{subequations}
}
where $\lambda$ denote a vector of Lagrangian multipliers associated with the linear constraints.

Taking into account that the problem is convex and only local nonlinearity exists in the complementary slackness component of the perturbed KKT conditions, we apply simple step-limiting only on dual variables $\mu$ (corresponding to inequalities) and $t$ to make each iteration update faster and more efficient. 
These heuristics were originally developed in our previous work \cite{SUGAR_SE_WLAV} for bus-branch models and this paper extends them to node-breaker models to solve the GSE problem. 

Algorithm \ref{alg:LP solver} illustrates our circuit-theoretic variable limiting heuristics.
{\color{black}In this algorithm, step 1 adjusts the update of $\mu$ based on the limits defined in the dual feasibility $\mu \succeq 0$ and the stationarity $\overline{\mu}+\underline{\mu} = c$. And step 2 adjusts $t_j$ to guarantee the satisfaction of the primal feasibility $|n|\preceq t$.}
\begin{algorithm}
	\caption{Variable limiting heuristics to solve LP problem}
	\label{alg:LP solver}
	\KwIn{previous solution $\mu_{old}$,
	new solution $\mu,t,n$, step limit $d$}
	\KwOut{new solution $\mu,t$ after limiting}
	For each element $\mu_j$ in $\mu$:
	$$\Delta\mu_j=\mu_j-\mu_{old,j}$$
	$$dir = sign(\Delta\mu_j)$$
	$$h = \begin{cases} c_j-\mu_{old,j} & dir \ge 0 \\ \mu_{old,j} & dir < 0 \end{cases}$$
    $$\mu_j=dir*\min(d,h)$$
	
	For each element $t_j$: 
	$$t_j = \begin{cases} 2|n_j| & |n_j|>t_j \\ t_j & else \end{cases}$$
\end{algorithm}

As \eqref{prob: differentiable ckt-GSE} is convex and applicable to realistic settings of meters (both SCADA meters and modern PMUs), the proposed method improves earlier works of WLAV-based AC-GSE \cite{TESE-GSE-PMUabur},\cite{convexTESE-SDP-weng} which were either limited to only PMUs \cite{TESE-GSE-PMUabur} or applied non-scalable relaxation techniques \cite{convexTESE-SDP-weng} to convexify the problem.

\subsection{Hypothesis test to validate wrong switch status} \label{sec:hypothesis_test}
By using WLAV formulation on the NB model, the ckt-GSE algorithm in Section \ref{sec:GSE formulation} provides a robust solution that implicitly rejects any data errors. While the sparsity of the noise vector is already indicative of the location of suspicious data samples, we propose the use of a hypothesis test to formally identify wrong switch statuses (i.e., topology errors). It follows from grid physics that an open switch should have zero current flow, whereas a closed switch should have nearly zero voltage across it, see Table \ref{tab:hypothesis test}. In this work, thresholds $\tau_I$ and $\tau_V$ are chosen from empirical values $\tau_I=0.01, \tau_V=0.01$. 
\begin{table}[htbp]
\caption{Hypothesis test to detect wrong switch status}
\label{tab:hypothesis test}
\begin{tabularx}{\linewidth}{lll}
\hline
\bf Measured status & \bf Hypothesis test & \bf Conclusion\\
\hline
Open & \begin{tabular}[c]{@{}l@{}}  $|I_{sw}|>\tau_I?$ \end{tabular} & \begin{tabular}[c]{@{}l@{}}If YES, switch should be closed\end{tabular}\\
\hline
Closed & \begin{tabular}[c]{@{}l@{}}  $|V_{sw}|>\tau_V?$ \end{tabular} & \begin{tabular}[c]{@{}l@{}}If YES, switch should be open\end{tabular}\\
\hline
\end{tabularx}
\end{table}

%% file: 040experiment.tex
To validate the efficacy of the proposed models and method, we design experiments to answer the following questions:
\begin{enumerate}
    \item \textbf{Robustness:} Is the method robust against bad data and topology error?
    \item \textbf{When does it fail:} 
    How does the (solution accuracy) performance change as the number of data error increase?
    \item \textbf{Scalability:} Is the method applicable to large networks?
\end{enumerate}
\noindent 

{\color{black}
\textbf{Reproducibility:} All test cases are from the CyPRES public dataset available at {\color{blue} \url{https://cypres.engr.tamu.edu/test-cases}}. 
And all experiments are run on a laptop computer with 11th Gen Intel(R) Core(TM) i7-1185G7 @ 3.00GHz   1.80 GHz processor and 32 GB RAM. 

\textbf{Assumption of meter placement: }Today's industrial practices and guidelines \cite{PMU_install_guideline} suggest the installation of PMUs at plants generating more than 100 MVA, large load buses, and grid control devices. Thus, in this paper, we assume the installation of PMUs on every generation bus and traditional SCADA meters (RTUs) on other injection buses without generators. We further assume line power flow measurements at randomly selected transmission lines that have an RTU located at either the from or to node.
}

\subsection{Robustness: WLAV outperforms WLS}
 \begin{figure}[b]
	\centering	\includegraphics[width=0.9\linewidth]{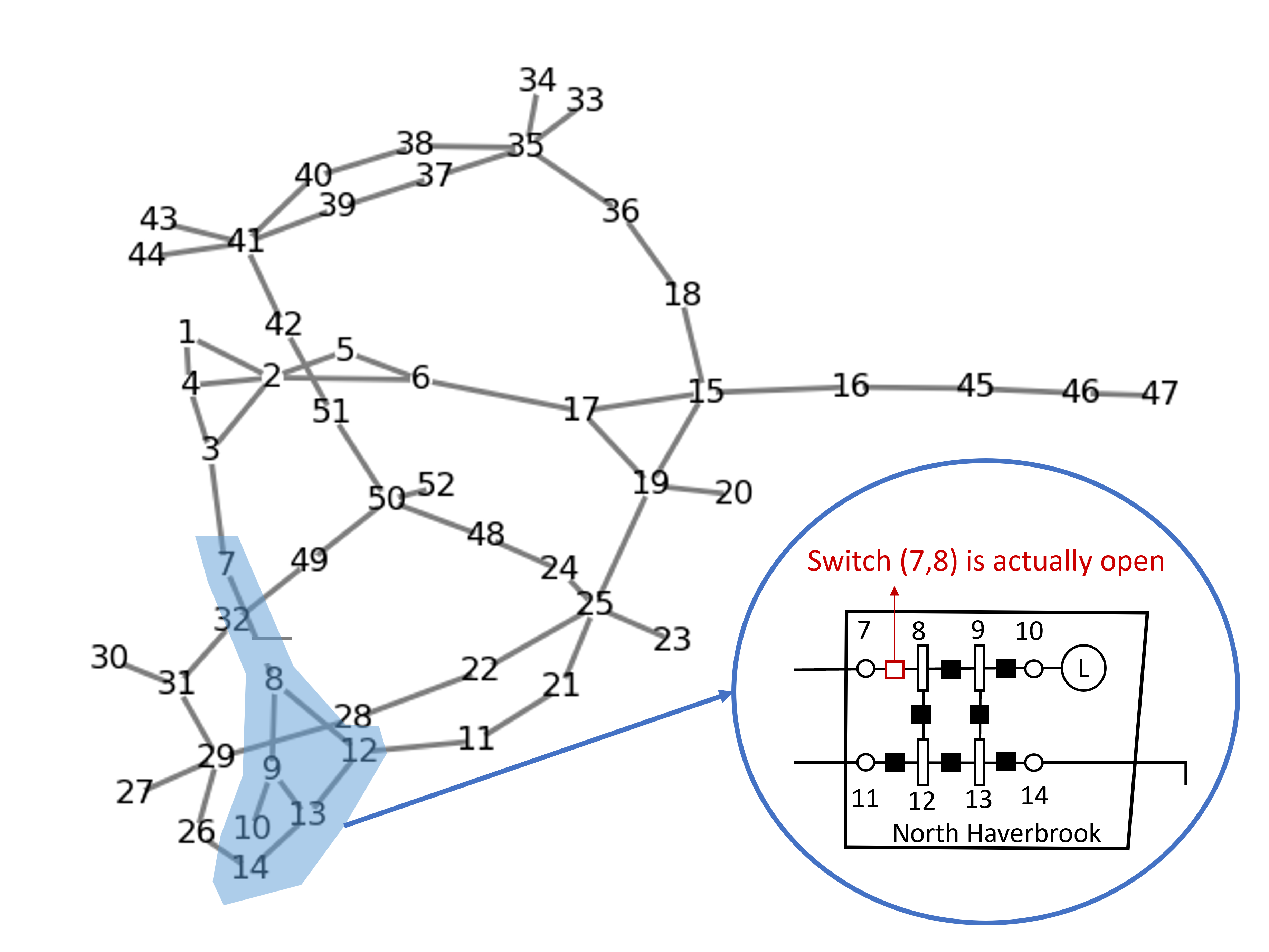}
	\caption[]{{\color{black} CyPRES 8 substation network. (The case is modified by opening the switch (7,8))}}
	\label{fig:8-substation case}
\end{figure}

{\color{black}Here, we evaluate the robustness of ckt-GSE method. Here the weighted least absolute value (WLAV) based robust estimator is expected to have two desirable properties that a weighted least square (WLS) method does not have:

\begin{itemize}
    \item  \textbf{automatically reject} data errors: the state solution is still accurate when data errors exist. In this paper, the evaluation metrics for solution accuracy include: 
    \begin{enumerate}
        \item root mean squared error (RMSE) which evaluates the overall deviation from the true states:
        \begin{equation}
            RMSE = \sqrt{||x_{est}-x_{true}||_2^2}
            \label{eq:rmse}
        \end{equation}
    \item number of inaccurate bus estimates: which is the number of buses whose estimated states have $>0.02 pu$ $|V|$ error or $>2^{\circ}$ phase angle ($\theta$) error, i.e.,
    \begin{align}
        &\text{Number of inaccurate bus estimates}\notag\\
        =&\sum_{busi} I\{|\Delta|V_i||>0.02, or|\Delta\theta_i|>2^\circ\}\label{eq:num of inaccurate bus}\\
        &\text{with } \Delta|V_i|=|V_i|_{est}-|V_i|_{true},\notag \\ &\Delta|\theta_i|=|\theta_i|_{est}-|\theta_i|_{true} \notag
    \end{align}
    A small value means that solution inaccuracy only exists
    regionally on a subset of buses.
    \end{enumerate}
    \item \textbf{identify} data errors: multiple types of data errors (even when they co-exist) which affect state estimation can be detected and localized:
    \begin{align}
        &\text{For PMU i, create alarm if } |n_{pmui}|>0.1\notag\\
        &\text{For RTU j, create alarm if } |n_{rtuj}|>0.1 \label{eq:bdd criteria}\\
        &\text{For sw k, raise suspicion if } |n_{swk}|>0.05,\notag\\
        &\text{ and create alarm by } \text{hypothesis test (Table \ref{tab:hypothesis test})}\notag
    \end{align}
     These bad data identification thresholds are empirically learned from our synthetic data, specifically by observing the data and finding a threshold value that effectively separates bad data points from normal ones, and they work well in our experiments. In real-world applications, the grid operators may need to learn their own optimal threshold from their real data, by observation, experience, or checking the area under curve (AUC) metric.
    However, due to the redundancy in realistic switch installation, some wrong switches will not affect the state estimation, and they are undetectable, as discussed later in Section \ref{sec:detectability}.  Thus in this work, we do not adopt any performance metric since they may not reflect the quality of estimation.  
\end{itemize}
}

Here, we consider the following types of data errors that can realistically occur and disrupt state estimation:
\begin{enumerate}
    \item \textbf{topology error}: either 1) a switch is actually \textit{open} but reported as \textit{closed},
    or 2) a switch is actually \textit{closed} but reported as \textit{open}
    \item \textbf{bad (continuous) data} from RTU or PMU, also known as \textit{(traditional) bad data}, which appears as a large deviation (1 p.u. in this paper) from the true value 
\end{enumerate}

We conduct experiments on an 8 substation node-breaker case.
Table \ref{tab:case info} and Fig. \ref{fig:8-substation case} show the case information and experiment settings.

\begin{table}[htbp]
\caption{Experiment settings on 8-substation case}
\label{tab:case info}
\begin{tabularx}{\linewidth}{m{0.16\linewidth}m{0.73\linewidth}}
\hline
{\bf Case name} & CyPRES 8-substation cyber-physical power system case \\
\hline
{\bf Case info} & 
\begin{itemize}
    \item 52 nodes, 49 breakers (switches)
    \item 5 generators (4 of them are active), 6 loads, 1 shunt 
    \item 1 transformer, 11 transmission lines
\end{itemize}\\
\hline
{\bf Synthetic meters: Location and Type} & \begin{itemize}
    \item \Statusmeass{} created on 49 breakers
    {\color{black}
    \item 5 PMU buses: each generator bus has a PMU installed to collect voltage and current phasors
    \item 7 RTU buses: each load bus has an RTU installed to collect $P_{rtu}, Q_{rtu}, |V|_{rtu}$ data
    }
    \item 22 RTU line meters (measurements include $P_{ij,rtu}, Q_{ij,rtu}$ and voltage magnitude at one end $|V_{i,rtu}|$ ) on selected lines
    \item Data generated by adding Gaussian noise (std=0.001) to power flow solution
\end{itemize}\\
\hline
{\color{black}
{\bf Data error generation}}
&
{\color{black} Bad RTU and topology errors are created (randomly):
 \begin{itemize}
     \item sw (7,8): actually open but measured as closed
     \item sw (20,19): actually closed but measured as open   
     \item bad RTU meter on bus 34: measurement of load values are perturbed by large random noise
 \end{itemize}
 }
\\
\hline
{\bf Hyper-parameters}
& 
\begin{itemize}
    \item{\color{black} RTU weights $= 1$, PMU weights $=1$}
    \item Switch weights $=0.001$
\end{itemize}
(See Section \ref{sec: sensitivity} for details of weight selection.)
\\
\hline
\end{tabularx}
\end{table}

\begin{figure*}[h]
     \centering
     \begin{subfigure}[b]{0.49\linewidth}
         \centering         \includegraphics[width=0.95\textwidth]{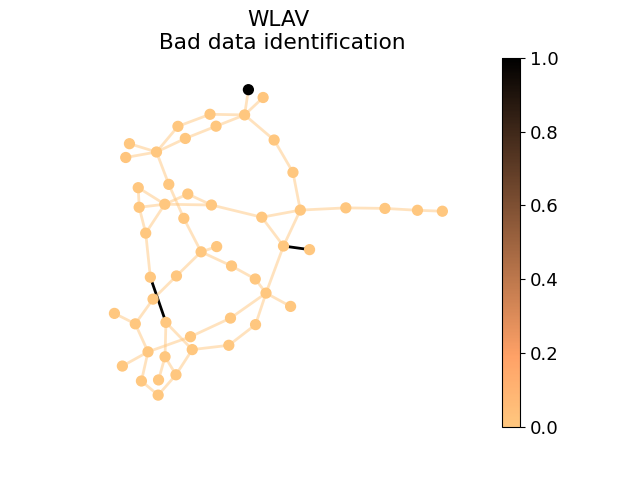}
         \caption{The WLAV-based ckt-GSE can identify 2 topology errors and 1 bad RTU. For better visualization, the node values $|n_{rtu}|, |n_{pmu}|$ are scaled to $[0,1]$ by $\frac{|n|}{max(n)}$; edge values are switch alarms (0 or 1).}
         \label{fig: wlav bdd graph}
     \end{subfigure}
     \hfill
     \begin{subfigure}[b]{0.49\linewidth}
         \centering         \includegraphics[width=0.95\textwidth]{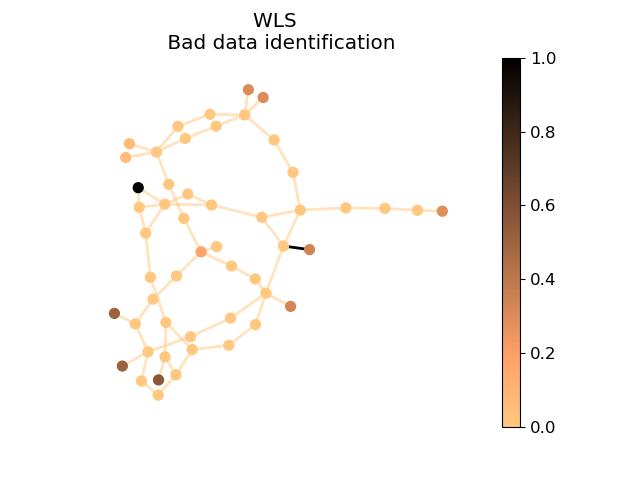}
         \caption{The WLS counterpart gives dense instead of sparse $n$ values. It only only recognizes 1 topology error, and  mistakenly indicate  bad data at many node locations.}
         \label{fig: wls bdd graph}
     \end{subfigure}
     \begin{subfigure}[b]{0.49\linewidth}
         \centering         \includegraphics[width=0.9\textwidth]{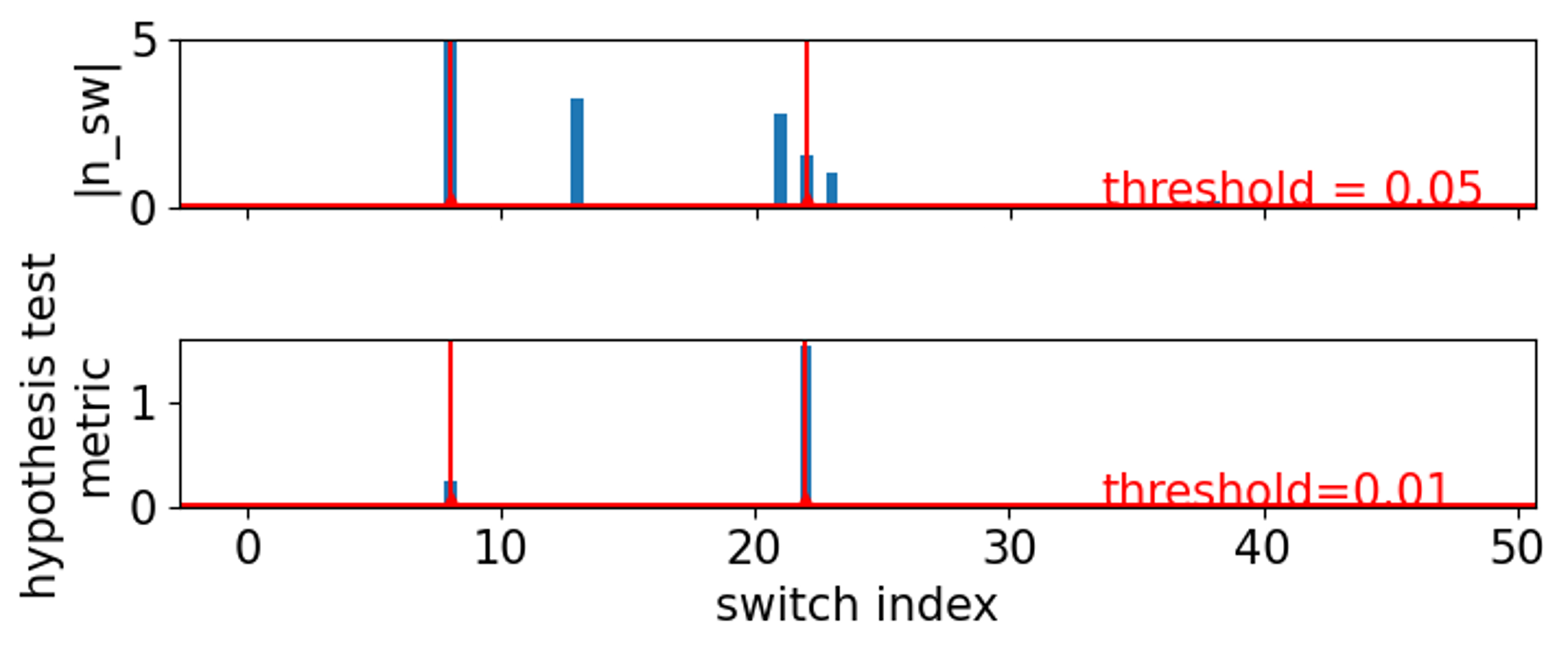}
         \caption{WLAV: sparse $n_{sw}$ 
         filters out suspicious switches and hypothesis test verifies the suspicion. All detection criteria are defined in (\ref{eq:bdd criteria}). Red vertical lines mark the true locations of topology errors.
          }
         \label{fig: wlav switch bdd}
     \end{subfigure}
     \hfill
     \begin{subfigure}[b]{0.49\linewidth}
         \centering         \includegraphics[width=0.9\textwidth]{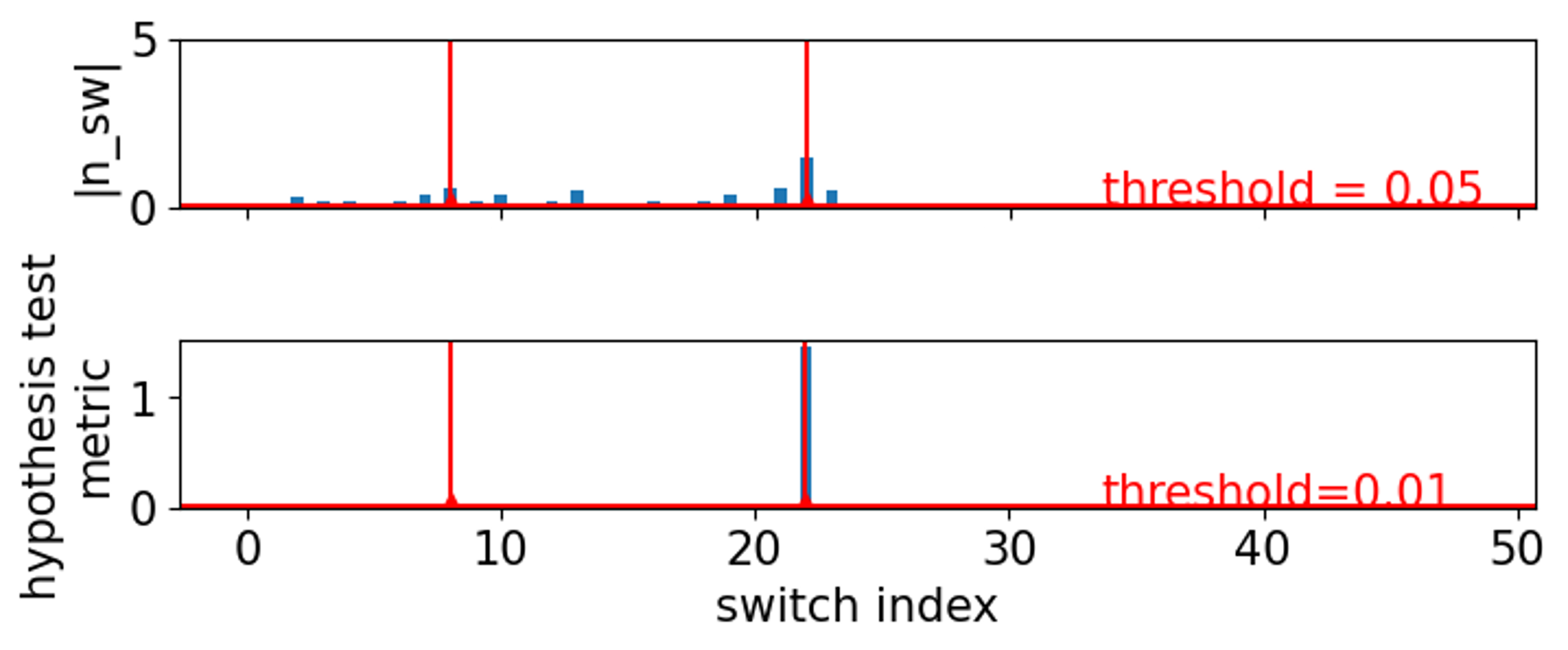}
         \caption{The WLS counterpart results in denser estimates of $n_{sw}$, i.e., more suspicious locations, but fails to recognize all topology errors after hypothesis test.}
         \label{fig: wls switch bdd}
     \end{subfigure}
     \begin{subfigure}[b]{0.49\linewidth}
         \centering         \includegraphics[width=0.95\textwidth]{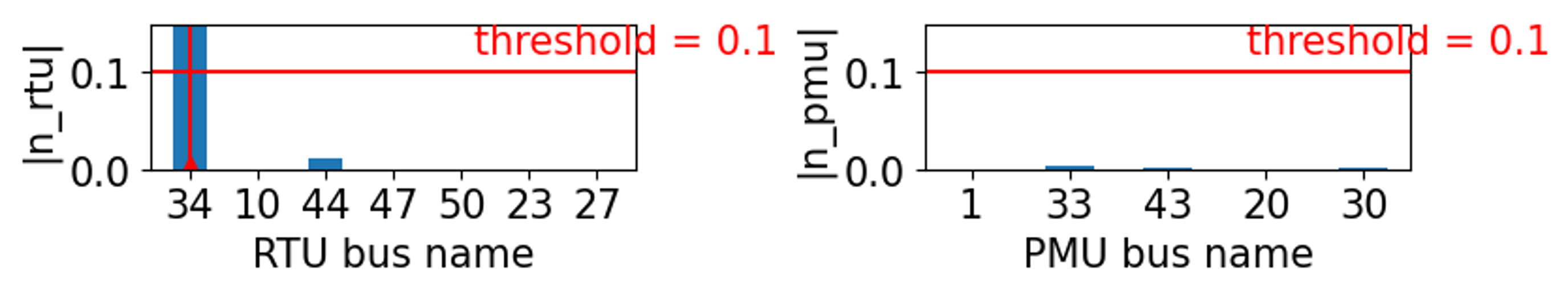}
         \caption{WLAV results in sparse $n_{rtu}$ to clearly identify the bad RTU at bus 34. (The red vertical line marks the true bad meter location.)}
         \label{fig: wlav bus bdd}
     \end{subfigure}
     \hfill
     \begin{subfigure}[b]{0.49\linewidth}
         \centering         \includegraphics[width=0.95\textwidth]{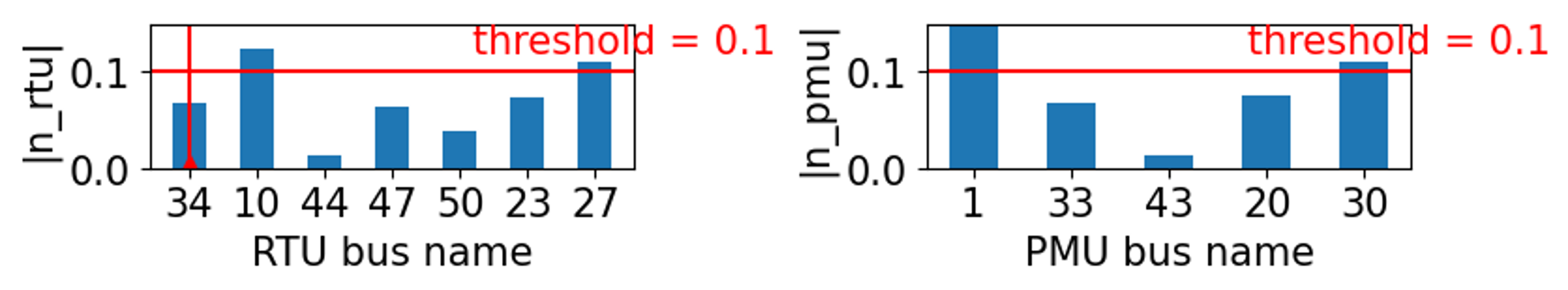}
         \caption{The WLS counterpart results in dense $n_{rtu}$ and $n_{pmu}$, leading to false bad data alarms. (Red vertical line marks the true bad meter.)}
         \label{fig: wls bus bdd}
     \end{subfigure}
     \begin{subfigure}[b]{0.49\linewidth}
         \centering         \includegraphics[width=0.9\textwidth]{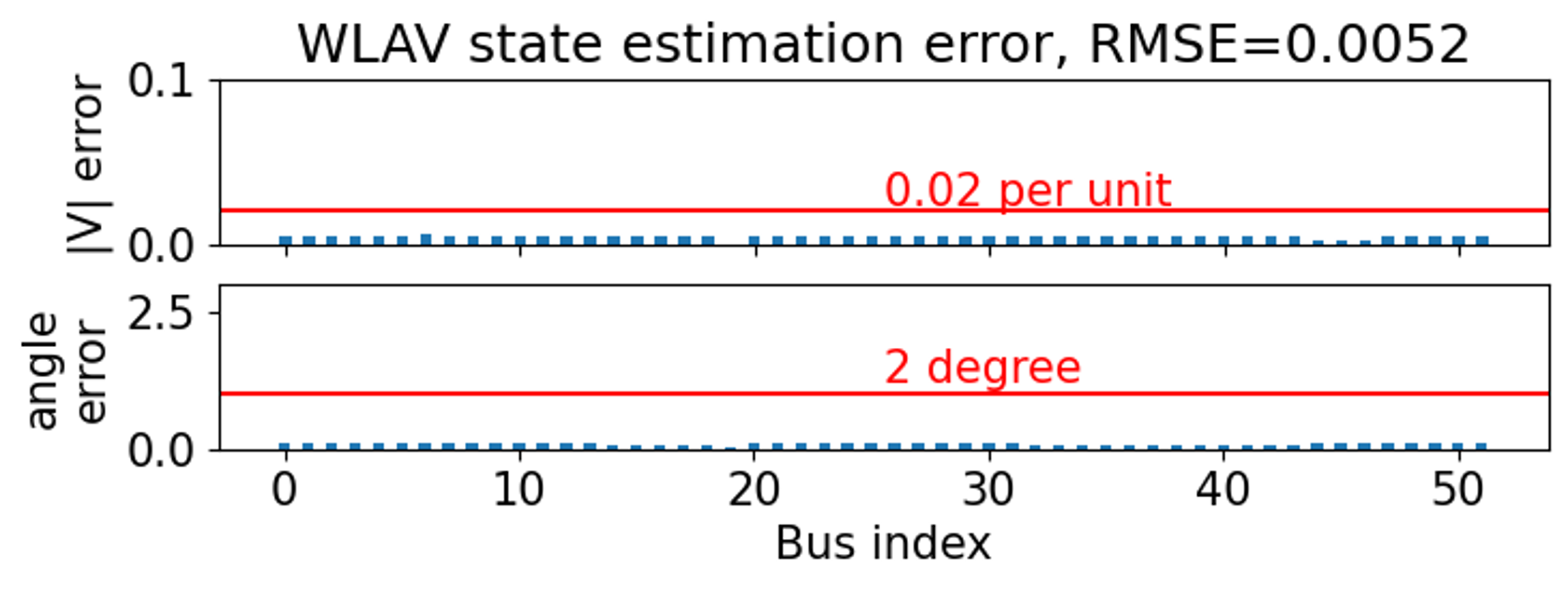}
         \caption{WLAV solution quality is not affected in the presence of bad-data. Bus voltages are still accurate.}
         \label{fig: wls voltage error}
     \end{subfigure}
     \hfill
     \begin{subfigure}[b]{0.49\linewidth}
         \centering         \includegraphics[width=0.9\textwidth]{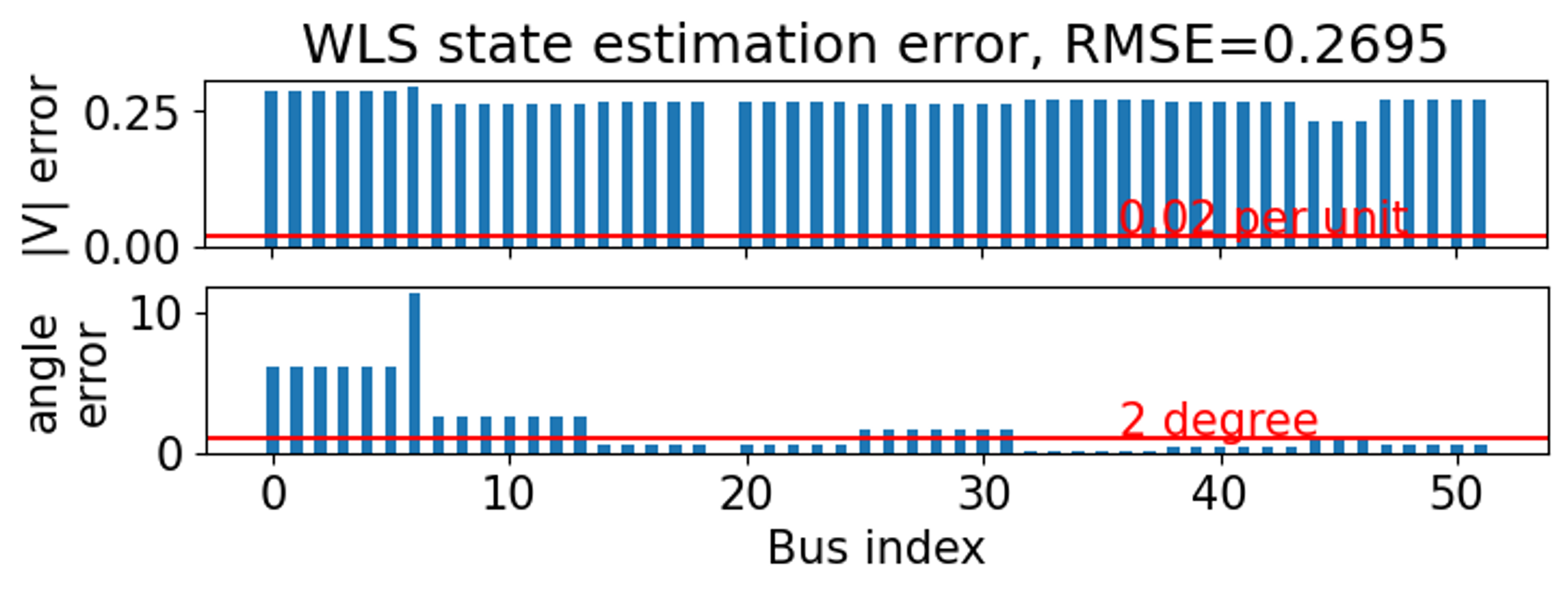}
         \caption{WLS solution is inaccurate: 51 nodes have $>0.02pu$ $|V|$ error, 14 nodes have $>2^\circ$ angle error. }
         \label{fig: wls voltage error}
     \end{subfigure}
        \caption{{\color{black} Robustness of WLAV (left) vs WLS (right): WLAV better identifies data errors and obtains accurate estimates. }}
        \label{fig: wlav vs wls}
        
\end{figure*}

{\color{black}
Results in Fig. \ref{fig: wlav vs wls} demonstrate the robustness of the proposed WLAV-based ckt-GSE model by comparing it against its WLS counterpart (the objective function minimizes the weight least squares of $n$ and the constraints remain the same). 
In terms of data error identification, Fig. (\ref{fig: wlav bdd graph}) and (\ref{fig: wls bdd graph}) demonstrate that the proposed WLAV model can provide sparse error indicators to precisely identify the topology errors and bad RTU bus; however, the WLS method fails to identify all topology errors and instead results in false alarms at many bus locations. 
Fig. (\ref{fig: wlav switch bdd}) - (\ref{fig: wls bus bdd}) further illustrates how the values of $n_{sw}, n_{rtu}, n_{pmu}$ along with hypothesis test can effectively identify different data errors. 
Further, in terms of the accuracy of state estimates, the WLAV model provides accurate solution with significantly smaller $|V|$ error, angle error, and RMSE. In contrast, the WLS solution  is significantly perturbed by data errors.

}

\begin{figure*}[!b]
     \centering
     \begin{subfigure}[b]{0.32\linewidth}
         \centering         \includegraphics[width=0.99\textwidth]{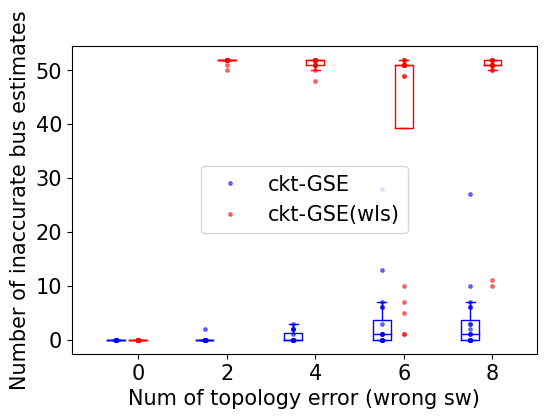}
     \end{subfigure}
     \hfill
     \begin{subfigure}[b]{0.32\linewidth}
         \centering         \includegraphics[width=0.99\textwidth]{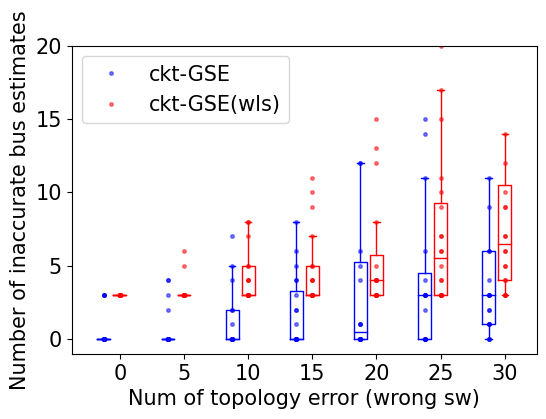}
     \end{subfigure}
     \begin{subfigure}[b]{0.32\linewidth}
         \centering         \includegraphics[width=0.99\textwidth]{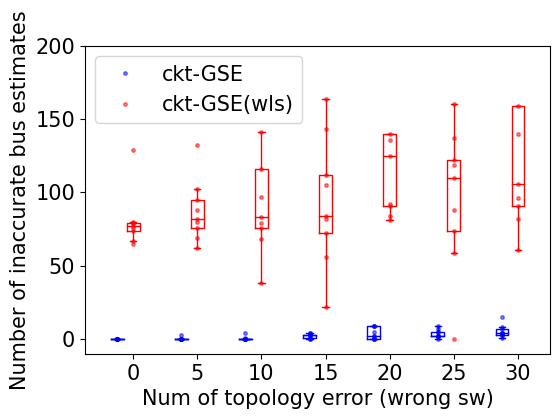}
     \end{subfigure}
     \hfill
     \begin{subfigure}[b]{0.32\linewidth}
         \centering         \includegraphics[width=0.99\textwidth]{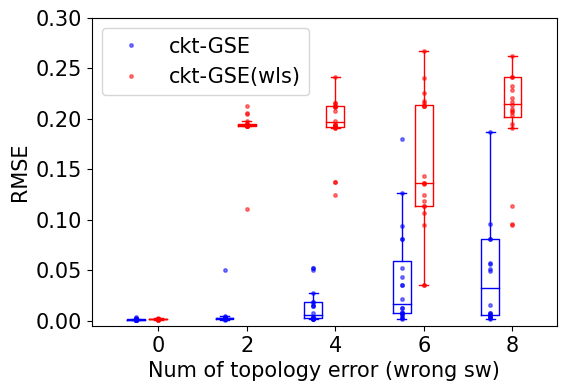}
         \caption{8-substation case: 51 nodes, 49 sw(itches)}
     \end{subfigure}
     \hfill
     \begin{subfigure}[b]{0.32\linewidth}
         \centering         \includegraphics[width=0.99\textwidth]{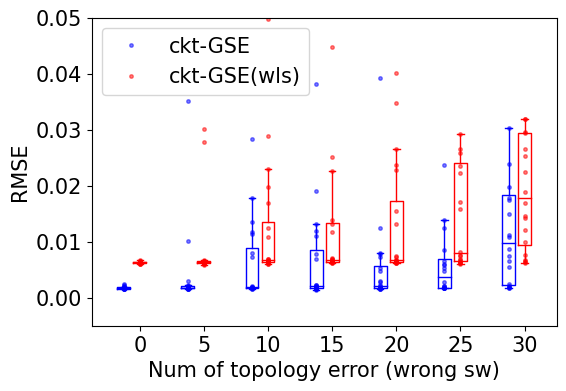}
         \caption{300-substation case: 1.6k nodes, 1.8k sw}
     \end{subfigure}
     \hfill
     \begin{subfigure}[b]{0.32\linewidth}
         \centering         \includegraphics[width=0.99\textwidth]{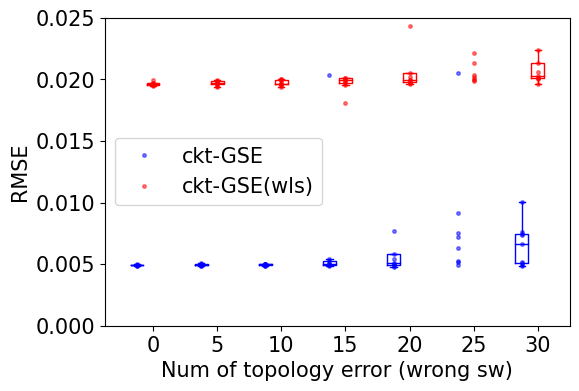}
        \caption{2000-substation case: 24k nodes, 23k sw}
     \end{subfigure}
    \caption{{\color{black} 
    Robustness on different sized networks: The top row shows the number of inaccurate bus estimates defined in (\ref{eq:num of inaccurate bus}) where a small value means inaccurate estimates only exist on a subset of nodes, and the bottom row shows RMSE defined in (\ref{eq:rmse}) which reflects the overall inaccuracy of solution. ckt-GSE remains very robust under low (sparse) penetration of topology errors and degrades robustly as topology errors grow, i.e., inaccuracy gradually appears on a larger subset of nodes. In contrast, the WLS model is not robust: even a few topology errors result in widespread and significant state estimate inaccuracies. }}
        \label{fig: robustness boundary}
        
\end{figure*}

\subsection{The boundary of robustness: when does it fail?}

{\color{black}
As the WLAV estimation algorithm achieves its desired robustness by enforcing sparsity, it relies on a basic assumption that the data errors are sparse. However, this property does not hold under higher penetration of data errors. In this Section, we explore how the growing percentage of topology errors will affect the robustness of the ckt-GSE and its WLS counterpart (for comparison). 

\begin{table}[htbp]
\caption{\color{black}Experiment settings on different cases}
\label{tab: exp setting, 3 cases}
\begin{tabularx}{\linewidth}{ll}
\hline
\bf CASE &\bf \begin{tabular}[c]{@{}l@{}} Settings\\(sec) \end{tabular} \\
\hline
\begin{tabular}[c]{@{}l@{}}
\textbf{8-substation case}\\
- 52 nodes\\ - 49 switches\\
\end{tabular}& \begin{tabular}[c]{@{}l@{}}
- 5 PMUs, 7 RTUs, 22 RTU line meters \\
- experiments repeated 20 times with \\different random data error locations
\end{tabular}\\
\hline
\begin{tabular}[c]{@{}l@{}}
\textbf{300-substation case}\\
- 1598 nodes\\ - 1816 switches\\
\end{tabular} & \begin{tabular}[c]{@{}l@{}}- 69 PMUs, 224 RTUs, 608 RTU line meters \\
- experiments repeated 20 times with \\different random data error locations\end{tabular}\\
\hline
\begin{tabular}[c]{@{}l@{}}
\textbf{Texas CP-2000 case}\\
\textbf{(2000-substation case)}\\
- 24360 nodes \\- 22632 switches\\
\end{tabular}  & \begin{tabular}[c]{@{}l@{}}- 522 PMUs, 1524 RTUs, 4690 RTU line meters \\
- experiments repeated 10 times with \\different random data error locations\end{tabular}\\
\hline
\end{tabularx}
\footnotesize{
*The Texas CP-2000 case is a cyber-physical model built from the footprint of the Texas grid.\\
*PMUs are placed on generation buses, RTUs are placed on load buses, and RTU line meters are placed on random lines.\\
}
\end{table}

Table \ref{tab: exp setting, 3 cases} shows the experiment settings for different cases and Fig. \ref{fig: robustness boundary} shows the results. We evaluate solution quality under a growing number of topology errors. 
Results show that
ckt-GSE has nearly zero inaccurate bus estimates and nearly zero RMSE under a small number of topology errors. This means ckt-GSE remains very robust under low (sparse) penetration of topology errors. As we have more topology errors, ckt-GSE degrades in a robust way: it makes mistakes at a subset of locations (regionally in subsets around the wrong switches), whereas the remaining bus locations still obtain accurate estimates.  As there are more topology errors, inaccuracy gradually spreads out. Whereas for the WLS counterpart, i.e., ckt-GSE(wls), there is always a larger number of buses whose state estimation is inaccurate, and a larger RMSE. This means the WLS solution is not robust and inaccuracy is wide-spread  even with a few topology errors.


\begin{table*}[b]
\caption{\color{black} Mathematical comparison of ckt-GSE against state-of-the-art methods.}
\begin{tabular}{l|l|l|l}
\hline
method                                                            & \multicolumn{1}{c|}{GSE-pmu\cite{TESE-GSE-PMUabur}}       & \multicolumn{1}{c|}{GSE-SDP\cite{convexTESE-SDP-weng}}    & \multicolumn{1}{c}{\begin{tabular}[c]{@{}c@{}}ckt-GSE (this work)\end{tabular}}     \\ \hline
\begin{tabular}[c]{@{}l@{}}data \\ assumption\end{tabular} & \begin{tabular}[c]{@{}l@{}} 
- PMUs placed to measure voltage, current\\and switch (or circuit breaker, CB) flows;\\
- system observable by PMUs {\color{red}(unrealistic)}.\end{tabular}       & \begin{tabular}[c]{@{}l@{}}
- (SCADA) RTU and PMU  placed to collect\\ power, voltage, current at buses, lines, switches;\\ - system observable by RTU and PMU data.\end{tabular}    & \begin{tabular}[c]{@{}l@{}}- RTU and PMU placed \\ - system observable by RTU, PMU and\\ zero-injection pseudo measurements\\
(less real data needed)\end{tabular}  \\ 
\hline
\begin{tabular}[c]{@{}l@{}}problem\\ formulation\end{tabular}     & \begin{tabular}[c]{@{}l@{}}
$\min||Wr||_1$\\ 
s.t. $ z = [D,M]\begin{bmatrix} x\\f\end{bmatrix} + r$
\\ $z$: PMU data of voltage, current, CB flows \\ $x$: AC states\\ $f$: (switch) breaker states\end{tabular}                & \begin{tabular}[c]{@{}l@{}}
$\min||Wr||_1$\\ 
s.t. $ z = h(x)+M f + r$
\\ $z$: RTU, PMU data of voltage, current, CB flows \\ $x$: AC states\\ $f$: (switch) breaker states\\
The nonlinear $h(x)$ comes from RTU data \end{tabular}             & \begin{tabular}[c]{@{}l@{}}
$\min||Wn||_1$\\ 
s.t. KCL constraints 
$ [Y,B]\begin{bmatrix} x\\n\end{bmatrix} = J$
\\
- PMU embedded in J, RTU embedded in Y.\\
- zero-injection (ZI) buses included in hard\\ constraints, used as error-free information. \\
\end{tabular}  \\
\hline
solver                              & \begin{tabular}[c]{@{}l@{}}- converted to differentiable form using\\ min-max model.\\ - solved with Simplex method.\\
\end{tabular} & \begin{tabular}[c]{@{}l@{}}- convexified via relaxation using SDP.\\ - solved by SDP solver.\\ \end{tabular} & \begin{tabular}[c]{@{}l@{}}- converted to differentiable form \\ using slack-t model (See math in (\ref{prob: differentiable ckt-GSE}))\\ - solved by circuit-based interior-point \\ solver\end{tabular} \\ 
\hline
weakness & 
\begin{tabular}[c]{@{}l@{}}
- On high-dimensional large cases, Simplex\\ faces numerical instability in pivoting \\operations and fails to converge.\\
(Fig. \ref{fig:scalability} shows the low speed of Simplex.)\\
- Undetectability issues.
\end{tabular}
&\begin{tabular}[c]{@{}l@{}}
- SDP is extremely difficult and insufficient\\ on large scale problems\cite{large-scale-SDP}; \\- SDP problem requires finding the rank-one \\solution, however, after relaxation, this is not\\ always possible, thus making solutions infeasible.\\
- Undetectability issues.
\end{tabular}
&
 \begin{tabular}[c]{@{}l@{}}
- Undetectability issues (see Section \ref{sec:detectability})\\ which is caused by redundant design\\ of power systems and unavoidable on\\
node-breaker models.
 \end{tabular}
\\
\hline
\end{tabular}
\label{tab: theoretical comparison}
\end{table*}

Thus the main finding is that the robustness of ckt-GSE degrades when the population of data errors becomes large. This holds for all WLAV-based models in general. The limitation is due to the violation of the sparse-data-error assumption and the algorithm's sensitivity to topology errors. Section \ref{sec: sensitivity} includes more discussions on the algorithm's sensitivity to different types of data errors.
}

\subsection{Scalability}
\begin{figure}[h]
	\centering
	\includegraphics[width=0.75\linewidth]{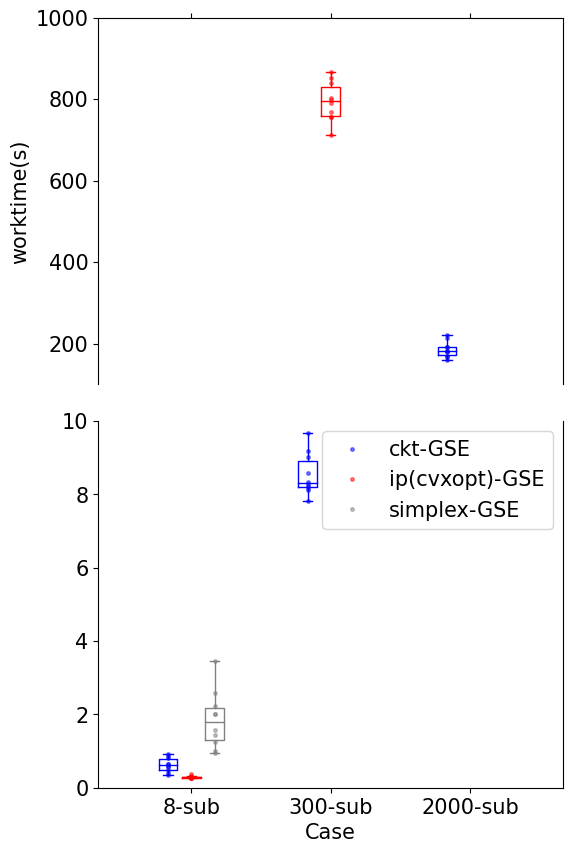}
	\caption[]{\color{black} Scalability: worktime of circuit-based (ckt) solver VS a standard interior-point (IP) solver in CVXOPT toolbox and a Simplex solver in Scipy toolbox. 
    Our ckt solver is efficient on different sized networks. CVXOPT is comparable with ckt solver only on small 8-substation case, whereas becomes significantly slower on larger cases and even fails on 2000-substation case (thus is not shown).
    Simplex solver is the slowest. It only works on the smallest case and fails on others (thus is not shown). } 
	\label{fig:scalability}
\end{figure}

The ckt-GSE method needs to be time-efficient on large-scale networks to be applicable in real-world control rooms.
{\color{black}Here we evaluate the speed of our proposed circuit-based (ckt) solvers by comparing with standard LP solvers:
\begin{itemize}
    \item interior-point (IP) solver in python CVXOPT toolbox
    \item Simplex method in SciPy which solves min-max model
\end{itemize}
Fig. \ref{fig:scalability} shows the speed performance on different sized networks. By comparison, our ckt solver is significantly faster than standard toolbox on large scale cases. 
}

%% file: 050discussion.tex
\section{Discussion}\label{sec: discussion}

{\color{black}
\subsection{ckt-GSE and state-of-the-arts}

To better clarify the advantage of ckt-GSE over the existing approaches, this Section discusses more on the two state-of-the-art models, GSE-pmu\cite{TESE-GSE-PMUabur} and GSE-SDP\cite{convexTESE-SDP-weng}. Table \ref{tab: theoretical comparison} compares them in measurement assumptions, mathematical formulation and problem solving strategies. The disadvantages of existing methods are demonstrated. 
}

\begin{figure*}[ht]
	\centering\includegraphics[width=0.9\linewidth]{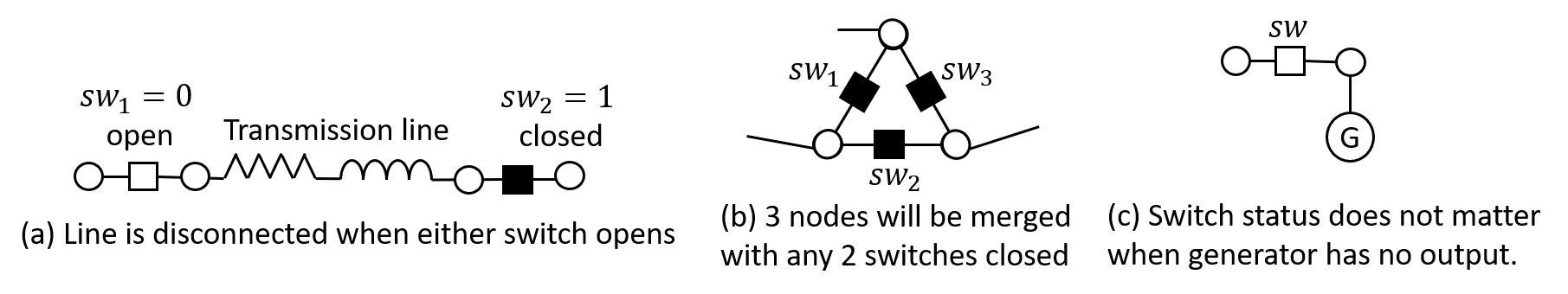}
	\caption[]{Detectability issues: three examples where the ckt-GSE algorithm will have problems identifying wrong \statusmeass{}. See Table \ref{tab:detectability issues} for illustration.} 
	\label{fig:detectability}
\end{figure*}

\begin{table*}[b]
\caption{Detectability issues of wrong status data on node breaker model}
\label{tab:detectability issues}
\begin{tabularx}{\linewidth}{m{0.19\linewidth}m{0.41\linewidth}m{0.33\linewidth}}
\hline
\bf Realistic condition & \bf Detectability issues & \bf Impact of the issue\\
\hline
\textbf{Equivalent line switches:} a realistic transmission line usually has switching devices at both ends of it, see Fig. \ref{fig:detectability}(a).
& \textit{Undetectability:} When one switch is open, any wrong status on the other switch is undetectable. This is because the transmission line is disconnected with either switch open, and the other switch, whether open or closed, does not impact the true grid states.
& Such instances of undetectability do not affect the quality of ckt-GSE solution as it has no impact on other grid states. From the viewpoint of  bus-branch model, such wrong switch status has no impact on grid topology.\\
& \textit{mis-localization}: When one switch is open and the other is closed, any wrong status on the open switch can be mis-localized at the wrong position. E.g., let $0$ denote open, and $1$ denote closed, when the true status is $[sw_1,sw_2]=[0,1]$, measured status is $[1,1]$, the wrong status localization may estimate the status to be $[1,0]$.
& Such mis-localization has no impact on the ckt-GSE solution since the true state and the mis-localized state are equivalent, with the same system topology in the bus-branch model.\\
\hline
\textbf{Cyclic connection of switches:} due to system redundancy some closed switches can form a cyclic graph, see Fig. \ref{fig:detectability}(b). 
&  \textit{Undetectability:} In Figure \ref{fig:detectability}(b), when any 2 switches are closed, the status of the third switch, whether closed or open, has no impact on the grid operation. Therefore, we cannot detect the wrong status indication for the third switch. 
&Such an instance of undetectability does not affect the quality of the ckt-GSE solution as it has no impact on other grid states. From the viewpoint of bus-branch model, topology remains unchanged independent of the status of the third switch. \\
& \textit{mis-localization}: When 2 switches are closed and one open, wrong status on any closed switch can be mis-localized. E.g., the true status is $[sw_1,sw_2,sw_3]=[1,1,0]$, the measured status is $[1,0,0]$, then a wrong estimation could be $[1,0,1]$
& Such mis-localization has no impact on the ckt-GSE solution as the mis-localization does not change the bus configuration in bus-branch model.\\
\hline
\textbf{Switching device connected to a node of degree one}, see Fig. \ref{fig:detectability}(c). 
&\textit{Undetectability:} In Figure \ref{fig:detectability}(c), when a generator has no output (produces no power), the switch status has no impact on the grid operating state and its bus-branch model. Therefore, the wrong status  on this switch is undetectable.
&The undetectability does not affect the quality of ckt-GSE as it has no impact on the grid operating state.\\
\hline
\end{tabularx}
\end{table*}

\subsection{Sensitivity issue: trade-off in weight selection}\label{sec: sensitivity}

As formulated in (\ref{eq:GSE}), each measurement device is assigned a weight in the objective function. This weight represents a level of confidence in each measurement and determines the algorithm’s sensitivity to different data errors. As our proposed method detects and localizes erroneous data by the sparse vector of noise terms $n$, the algorithm's sensitivity to data errors can be mathematically defined as the sensitivity of $n$ for any perturbation on the data (i.e., true data errors). 
Specifically, a lower weight $c_j$ for a particular measurement $j$ indicates the measurement is less trustworthy, and while minimizing $c_j|n_j|$ in the WLAV objective, a low $c_j$ tends to push the corresponding $n_j$ to a larger value, making the corresponding data error, if any, easily detectable.
Therefore, a lower weight makes the algorithm more sensitive to data errors at this location. This is a desirable feature as we expect less trustworthy meters to be more prone to gross data errors.

The selection of weights for continuous measurements $(\alpha, \beta, \gamma$ in \eqref{eq:GSE}) is statistically related to the variance (or dispersion) of the measurement tolerance (especially when assuming noise $n$ as Gaussian). Most existing works \cite{traditional-WLS-SE}\cite{TE-WLSE-BDI} set weights as $\frac{1}{\sigma^2}$ which is the reciprocal of the variance of the noise. This results in a statistical property wherein minimizing weighted least squares of the noise in the objective is equivalent to a maximal likelihood estimation (MLE) if we assume that noise $n$ follows Gaussian distribution. However, this can also lead to numerical issues as a high-quality measurement device (which has a very low noise variance) corresponds to an extremely high weight value which can cause ill-conditioning issues.
In this paper, to avoid extremely high weights, we scale the reciprocal of variance such that any RTU device with noise $\sigma=0.001$ has weight=1.

In contrast, the selection of weights for switches ($w$ in \eqref{eq:GSE}) requires additional tuning. 
Unlike continuous measurements, the switch statuses are discrete data, and the assumption of Gaussian noise no longer holds, making the statistical variance inapplicable. Instead, this paper’s selection of switch weights is based on considering a trade-off between convergence stability and the algorithm's sensitivity to topology error.
Specifically, when the weights of switches are high, the resulting low sensitivity to topology errors can cause a wrong switch status to be falsely identified as multiple bad continuous data and degrade ckt-GSE’s solution quality. While very low switch weights will result in a high sensitivity to topology errors and allow easy detection of wrong switch status data, low weights can cause numerical difficulties, which will deteriorate convergence efficiency of ckt-GSE. This paper applies hyper-parameter tuning to select the weights that give the lowest misclassification rate. Based on our empirical findings, the weights of switching devices should be lower than continuous meters to provide the necessary sensitivity for the wrong switch status. In this paper, we set all switch weights to $0.001$ for the 8 substation case, and $0.01$ for the 300-substation case and Taxas CP-2000 case.

 


\subsection{Undetectability issues: inevitable on node-breaker model}\label{sec:detectability}

Despite the ability of the NB model-based ckt-GSE to consider all switching devices and detect topology errors, not all wrong switch statuses are detectable (i.e., undetectability) using the observed data. In the real world, there exist cases where different grid configurations and anomaly scenarios have the same physical effect, and thus at times, one cannot accurately localize the source of an anomaly (i.e., mis-localization). These issues are limitations for all node-breaker based estimation methods such that some data errors can be \textit{undetectable} or \textit{mis-localized}, unless additional sources of information are included. The major cause is the redundancy of power system components. Fig. \ref{fig:detectability} illustrates 3 realistic scenarios where we may observe these issues, and Table \ref{tab:detectability issues} further describes the causes of potential \textit{undetectability} and \textit{mis-localization} of wrong switch status in these scenarios, as well as how these limitations affect the solution quality of ckt-GSE with node-breaker model.

%% file: 060conclusion.tex
This paper presents an AC-constrained generalized state estimation method built on a circuit-theoretic NB foundation (ckt-GSE). The method is robust, practical, and scales to large-scale networks subject to realistic grid data. Specific features and benefits of our ckt-GSE approach are:

\begin{itemize}
    \item \textbf{applicability to realistic data settings}: the device modeling considers both traditional (SCADA) RTU meters and PMUs, which is consistent with the realistic settings of meter installation and data collection in today’s power grid
    \item \textbf{robustness}: the formulation enables identifying the data error from the sparse slack variables $n$ and hypothesis test, while ensuring that estimates of switch status and grid states remain accurate and immune to data errors
    \item \textbf{convexity without relaxation requirement}: the linear device models for measurements and grid equipment result in convex (affine) constraints, and a resulting LP problem
    \item \textbf{scalability} of the formulation to large-scale networks (with more than 20k nodes in our experiments) with the use of circuit-theoretic heuristics
\end{itemize}

\section{Acknowledgement}
Work in this paper is supported in part by C3.ai Inc. and
Microsoft Corporation

%% file: appendix.tex
{\color{black}

\section{Appendix}

\subsection{How RTU model transforms non-linearity}
\label{apdx: RTU model} 

\noindent Here, we explain how circuit modeling transforms nonlinear measurement relationships in a physically meaningful way for convex and linear constraint formulation. Taking an example of an RTU power injection measurement at a bus, i.e., given $P_{rtu}$, $Q_{rtu}$ and $|V|_{rtu}$.
\begin{figure}[h]
	\centering
	\includegraphics[width=0.5\linewidth]{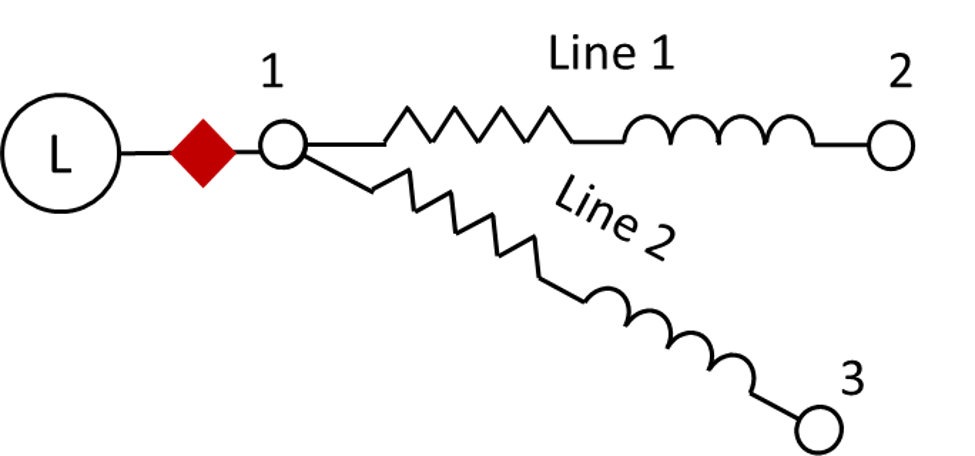}
	\caption[]{Suppose the power injection at a load bus is measured by an RTU device (in purple); the measurements are $P_{rtu}$, $Q_{rtu}$ and $|V|_{rtu}$.}
	\label{fig:explain rtu}
\end{figure}

In traditional modeling, these power injection measurements result in nonlinear models under polar coordinate, with voltage magnitude $|V|$ and phase angle $|\theta|$ being the state variables:
$$
P_{rtu} = \sum_{k\in{2,3}} |V_1||V_k|(G_{line(1,k)} cos\theta_{1k}+B_{line(1,k)} sin\theta_{1k}) + n_P
$$
 $$
 Q_{rtu} = \sum_{k\in{2,3}} |V_1||V_k|(G_{line(1,k)} sin\theta_{1k}-B_{line(1,k)} cos\theta_{1k}) + n_Q
$$
$$|V|_{rtu} = |V_1| + n_V$$

\noindent Now, with circuit modeling, we transform the original measurements $P_{rtu}$ and $Q_{rtu}$ to admittance measurements: 
$$G_{rtu}=\frac{P_{rtu}}{|V|_{rtu}^2}; B_{rtu}=-\frac{Q_{rtu}}{|V|_{rtu}^2}$$ 

\noindent These $G_{rtu},B_{rtu}$ can form linear constraints of the equivalent circuit model which are characterized by KCL equations under rectangular coordinate with real and imaginary voltage $V^R, V^I$ as state variables:
\begin{equation}
\begin{split}
  G_{rtu1} V_2^R-B_{rtu1}V_2^I+n_{rtu1}^R +\\G_{line1}(V_1^R-V_{2}^R) - B_{line2}(V_1^I-V_{2}^I)+\\G_{line2}(V_1^R-V_{3}^R) - B_{line2}(V_1^I-V_{3}^I)= 0  
\end{split}
\end{equation}

\begin{equation}
\begin{split}
 G_{rtu1}V_2^I+B_{rtu1}V_2^R+n_{rtu1}^I +\\ G_{line1}(V_1^I-V_{2}^I) + B_{line1}(V_1^R-V_{2}^R) +\\ G_{line2}(V_1^I-V_{3}^I) + B_{line2}(V_1^R-V_{3}^R)=0
\end{split}
\end{equation}

With linear constraints, the ckt-GSE problem is a linear programming problem with desirable properties stemming from: 
\begin{itemize}
    \item the use of rectangular coordinates results in linear transmission line models.
    \item the transformation of $P_{rtu}$ and $Q_{rtu}$ into $G_{rtu}$ and $B_{rtu}$ which leads to linear I-V relationships (KCL equations).
    \item formulating a constrained optimization problem under rectangular coordinate, thus having linear KCL constraints of the circuit model in the equality constraint set (instead of the commonly used unconstrained problem using the traditional "measurement = function + noise" measurement model, i.e., $z=h(x)+noise$)
\end{itemize}

}

%% file: biography.tex
\begin{IEEEbiography}
[{\includegraphics[width=1in,height=1.25in,clip,keepaspectratio]{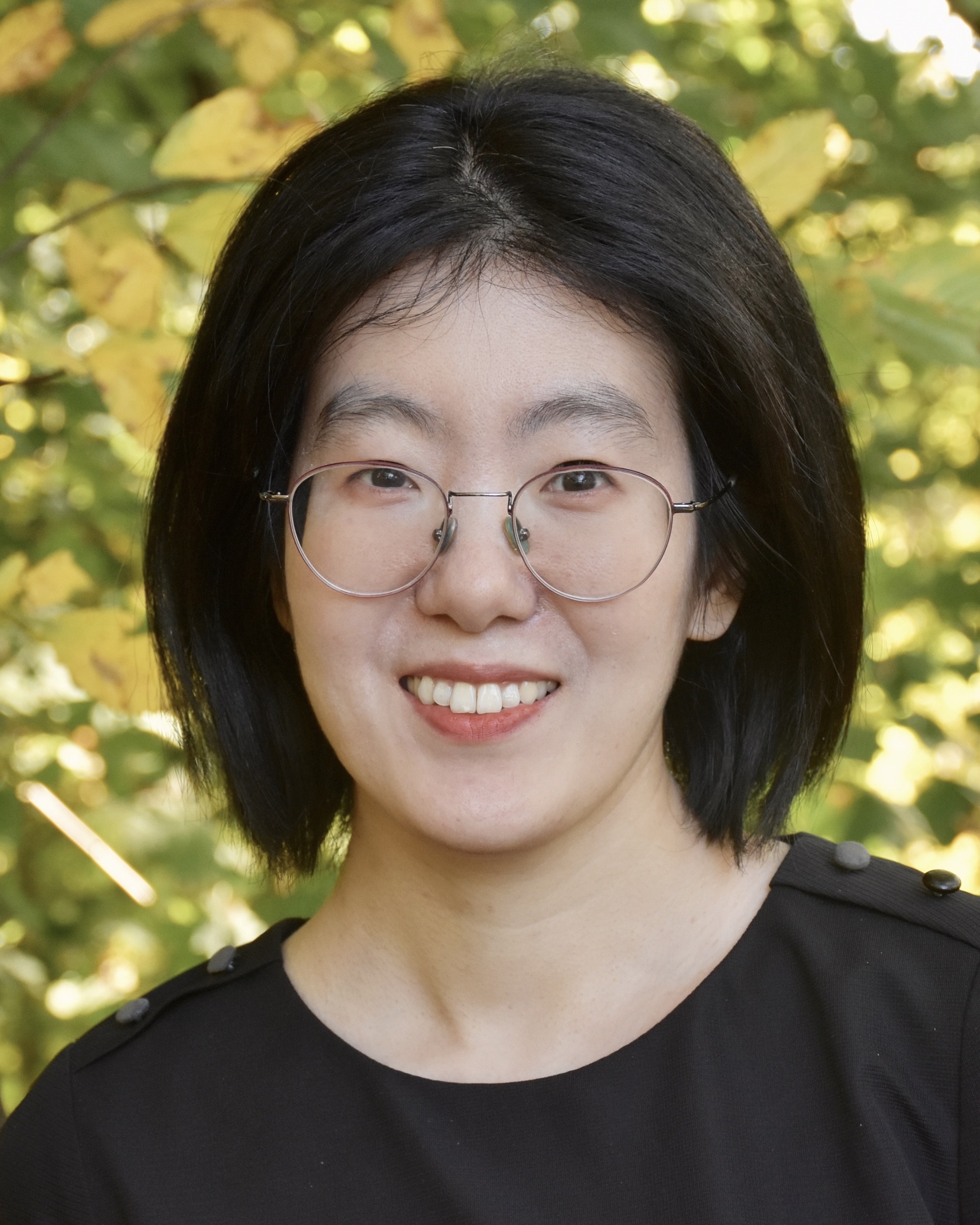}}] 
{Shimiao Li} She is a Ph.D. candidate in the Department of Electrical and Computer Engineering (ECE) at Carnegie
Mellon University, where she is advised by Professor Lawrence Pileggi. Previously, she received her B.E. degree in electrical engineering
from Tianjin University, China, in 2018.  Her research interests span the areas of operation and planning of real-world large-scale power grid, using
optimization, statistical and probabilistic models. Her previous work has mainly focused on scalable and robust algorithms
for situation awareness which includes system identification, anomaly detection, and root cause diagnosis
capabilities to gain perception of the present and future system conditions. 
\end{IEEEbiography}

\begin{IEEEbiography}
[{\includegraphics[width=1in,height=1.25in,clip,keepaspectratio]{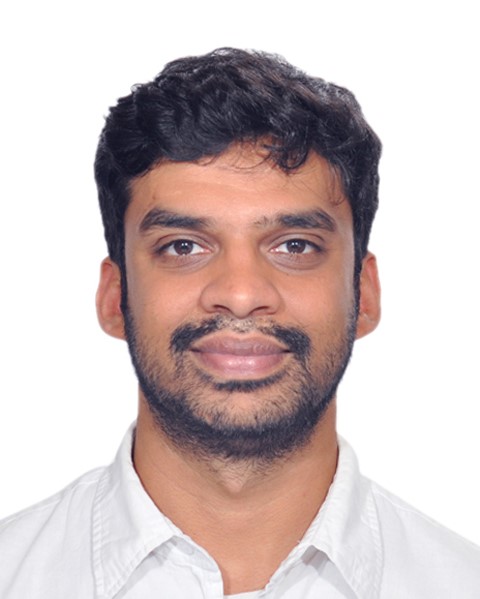}}] 
{Amritanshu Pandey} is an Assistant Professor in the Electrical and Biomedical Department at the University of Vermont with a courtesy appointment in the Electric and Computer Engineering and Engineering and Public Policy departments at Carnegie Mellon University. His overarching research goal is to develop electric energy system technologies to help combat climate change while modernizing the underlying system. In particular, he focuses on developing computational methods that address problems in the space of large-scale grid simulation and optimization, grid cybersecurity, and energy inequity. In the past, he worked on a novel circuit-theoretic simulation and optimization framework for power grids, culminating in a grid analytics tool: Simulation of Unified Grid Analysis and Renewables (SUGAR). He has won several best paper awards, including two best-of-the-best paper awards at the IEEE PES General Meeting in 2017 and 2021. He actively advises Pearl Street Technology, Inc. (PST) and has worked at MPR Associates, Inc. and PST.
\end{IEEEbiography}

\begin{IEEEbiography}
[{\includegraphics[width=1in,height=1in,clip,keepaspectratio]{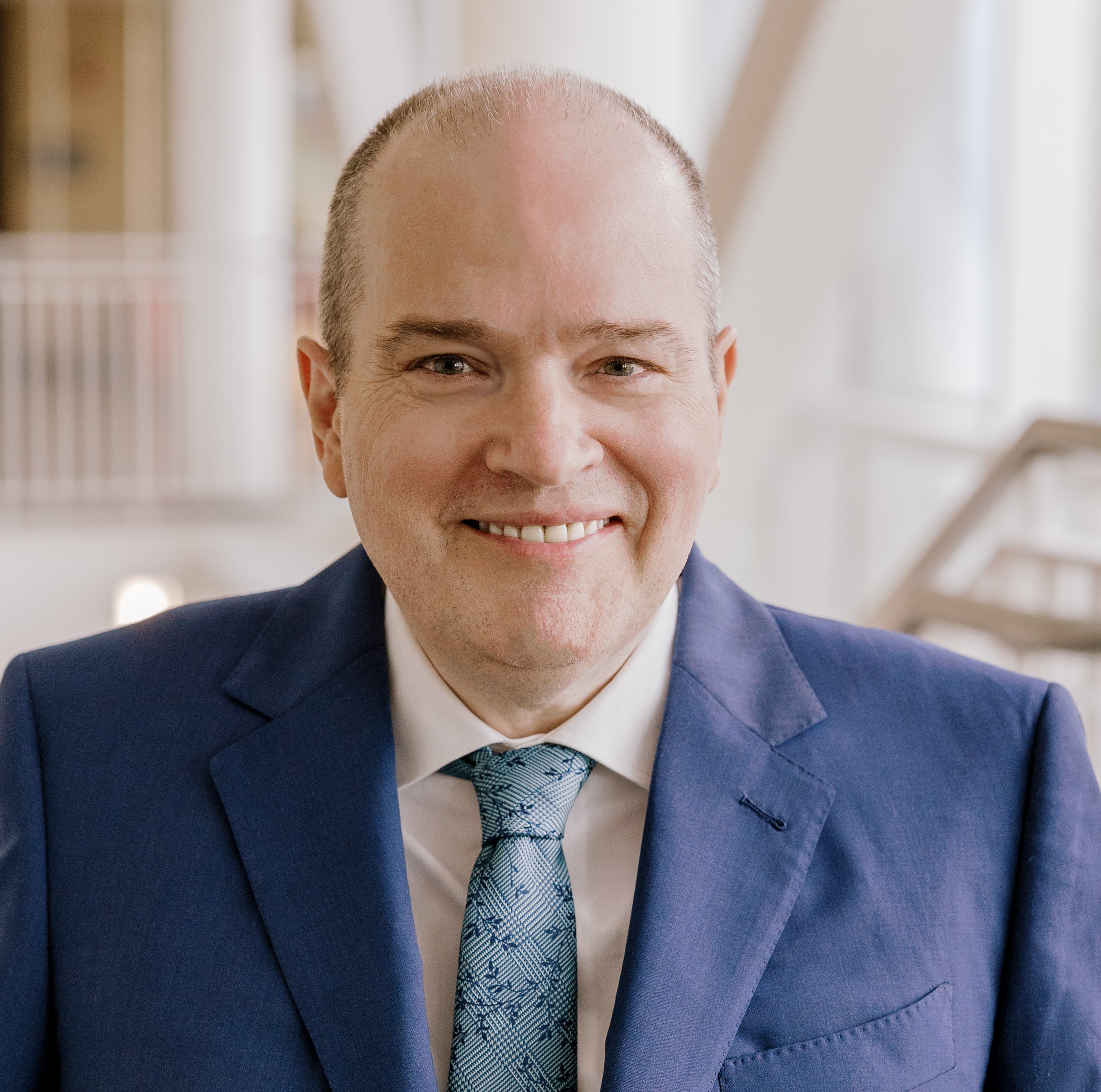}}] 
{Lawrence Pileggi} is the Tanoto professor and Head of electrical and computer engineering at Carnegie Mellon University, and has previously held positions at Westinghouse Research and Development and the University of Texas at Austin. He received his Ph.D. in Electrical and Computer Engineering from Carnegie Mellon University in 1989. His research interests include various aspects of digital and analog integrated circuit design, and simulation, optimization and modeling of electric power systems. He was a co-founder of  Fabbrix Inc., Extreme DA, and Pearl Street Technologies. He has received various awards, including Westinghouse corporation’s highest engineering achievement award, the Semiconductor Research Corporation (SRC) Technical Excellence Awards in 1991 and 1999, the FCRP inaugural Richard A. Newton GSRC Industrial Impact Award, the SRC Aristotle award in 2008, the 2010 IEEE Circuits and Systems Society Mac Van Valkenburg Award, the ACM/IEEE A. Richard Newton Technical Impact Award in Electronic Design Automation in 2011, the Carnegie Institute of Technology B.R. Teare Teaching Award for 2013, the 2015 Semiconductor Industry Association (SIA) University Researcher Award, and the 2023 Phil Kaufman Award from the Electronic System Design Alliance and the IEEE Council on Electronic Design Automation. He is a co-author of "Electronic Circuit and System Simulation Methods," McGraw-Hill, 1995 and "IC Interconnect Analysis," Kluwer, 2002. He has published over 400 conference and journal papers and holds 41 U.S. patents. He is a fellow of IEEE and a fellow of the National Academy of Inventors (NAI).
\end{IEEEbiography}